\documentclass[showpacs,superscriptaddress,aip,preprint]{revtex4-1}
\usepackage{palatino}
\usepackage{epsf}
\usepackage{amsmath,amssymb}
\usepackage{latexsym}
\usepackage{calc}
\usepackage{color}
\usepackage{graphicx}
\usepackage{float}

\usepackage{CJK}

\DeclareMathAlphabet\mathbfcal{OMS}{cmsy}{b}{n}

\begin{document}

\begin{CJK*}{GB}{} 
\title{Quantum-classical nonadiabatic dynamics of Floquet driven systems}
\author{Marco Schir\`o}
\affiliation{JEIP, USR 3573 CNRS, Coll\`ege de France, PSL Research University, 11 Place Marcelin Berthelot, 75321 Paris Cedex 05, France.$^{\star}$}
\author{Florian G. Eich}
\affiliation{HQS Quantum Simulations GmbH, Haid-und-Neu-Stra{\ss}e 7, D-76131 Karlsruhe, Germany}
\author{Federica Agostini~$^{\dagger}$}
\affiliation{Universit\'e Paris-Saclay, CNRS, Institut de Chimie Physique UMR8000, 91405, Orsay, France}
\begin{abstract}
We develop a trajectory-based approach for excited-state molecular dynamics simulations of systems subject to an external periodic drive. We combine the exact-factorization formalism, allowing to treat electron-nuclear systems in nonadiabatic regimes, with the Floquet formalism for time-periodic processes. The theory is developed starting with the molecular time-dependent Schr\"odinger equation with inclusion of an external periodic drive that couples to the system dipole moment. With the support of the Floquet formalism, quantum dynamics is approximated by combining classical-like, trajectory-based, nuclear evolution with electronic dynamics represented in the Floquet basis. The resulting algorithm, which is an extension of the coupled-trajectory mixed quantum-classical scheme for periodically driven systems, is applied to a model study, exactly solvable, with different field intensities.
\end{abstract}
\maketitle
\end{CJK*}

\small
\noindent
$^\dagger$~federica.agostini@universite-paris-saclay.fr\\
$^\star$~On leave from: Institut de Physique Th\'eorique, Universit\'e Paris-Saclay, CNRS, CEA, F-91191 Gif-sur-Yvette, France

\section{Introduction}
The development of theoretical approaches to treat light-matter interactions is nowadays a very active and productive field in various domains, from chemical physics to condensed-matter physics, from biophysics to material sciences. The most innovative studies are perhaps phenomena exhibiting strong coupling between the radiation field and matter, and requiring special frameworks even beyond the standard classical description of light~\cite{Ebbesen_ACR2016, Rubio_PRA2018, Rubio_NRC2018, Tokatly_EPJB2018, Maitra_EPJB2018, Rubio_JCTC2017, Maitra_PRL2019}. In other situations, for instance when the external field initiates and drives the highly non-equilibrium dynamics of molecular systems or solids, it is necessary to develop non-perturbative approaches to adequately describe light-matter interactions~\cite{Nitzan_JPCM2017, Rubio_JPB2020, Yang_JPB1995, Posthumus_RPP2004, Maitra_PCCP2017, Maitra_PRL2015, Suzuki_PRA2014, Nauts_JCP2007}. 

In photochemistry, the coupling of a molecular system with light is routinely used to steer or to drive chemical reactions~\cite{Banares_PCCP2015}. Molecular dynamics simulations of photochemical reactions, however, not always consider the external field explicitly in the calculations, even though some examples have been reported~\cite{Gonzelez_JCTC2011, Martinez_JCP2016, Curchod_JPCA2019, Vrakking_PCCP2011, Gonzalez_FD2011}. In condensed matter physics an emerging frontier is the optical control of material properties~\cite{Oka_AnnRevCondMat2018} achieved by shining strong and frequency selective light fields on solid state materials~\cite{Nicoletti_AdvOpt2016} and leading to a variety of light-induced collective phenomena, from transient superconductivity~\cite{Fausti_Science2011} to non-trivial topological phases~\cite{McIver_NatPhys2020}.

In recent years, in order to simulate and interpret these phenomena, some interest has been devoted to the integration of Floquet theory for time-periodic Hamiltonians with techniques for excited-state, nonadiabatic dynamics. Floquet treatment of the periodicity induced on a quantum system by an external field effectively maps the time-dependent Schrodinger equation into an eigenvalue problem for Floquet states and their associated quasi-energies, analogously to the Bloch theorem for electrons in periodic potentials~\cite{Floquet_1880, Sambe_PRA1973, Hanggi_PhysRep98}.
In the \textsl{Floquet picture}, time-dependent processes with periodic drive are analyzed in an extended vector space of ``physical'' electronic states and harmonics of the driving frequency. Therefore, electronic states can be interpreted as \textsl{dressed} states by the drive harmonics. The concept of static Floquet potential energy surfaces is, thus, introduced, which clearly generalizes the concept of static adiabatic, or Born-Oppenheimer, potential energy surfaces that are routinely used for interpreting field-free nonadiabatic processes~\cite{Curchod_WIRES2019}. Such a Floquet picture has been often employed in strong-field molecular physics, for interpretation purposes of bond-softening~\cite{Schumacher_PRL1990} and bond-hardening~\cite{Mies_PRL1992} processes, photo-dissociation~\cite{Mies_PRL1990, Taday_JPB2000, GiustiSuzor_PRA1988, Atabek_PRA1992}, dynamical alignment~\cite{Schmidt_PRA2005} and anti-alignment~\cite{Langley_PRL2001}. Recently, and due to the analogy between Born-Oppenheimer states/surfaces and Floquet dressed states/surfaces, the Floquet picture has been employed for actual simulations of driven time-dependent processes in molecules~\cite{Schmidt_PRA2017, Welsch_PRA2020, Subotnik_JCTC2020, Shalashilin_CP2018, Schmidt_PRA2016, Gonzalez_JPCA2012}. In particular, trajectory-based schemes for excited-state dynamics, where the concept of electronic potential energy surfaces that guide nuclear dynamics is of utmost importance, lend themselves, naturally, to be combined with the Floquet picture.

In this paper, we focus on the combination of the exact factorization of the electronic-nuclear wavefunction~\cite{EF_bookchapter_2020, Gross_PRL2010, Curchod_WIRES2019, Gross_TDDFTbook2018} with the Floquet formalism~\cite{Fiedlschuster_PhD2018, Schmidt_PRA2017}, devoting particular attention to the coupled-trajectory mixed quantum-classical (CT-MQC) algorithm~\cite{Gross_PRL2015, Gross_JCTC2016, Gross_JPCL2017, Maitra_JCTC2018, Tavernelli_EPJB2018, Agostini_EPJB2018}. CT-MQC is the numerical scheme allowing to solve the exact-factorization equations based on the quantum-classical approximation~\cite{Gross_EPL2014, Gross_JCP2014, Ciccotti_EPJB2018, Ciccotti_JPCA2020} of the nonadiabatic electron-nuclear problem, by introducing a trajectory-based solution of nuclear dynamics as formulated within the exact factorization. In the exact factorization, the nuclear wavefunction evolves according to a standard time-dependent Schr\"odinger equation where the dynamic, fully nonadiabatic, effect of the electrons is represented by a time-dependent vector potential and a time-dependent potential energy surface (TDPES). It has been shown~\cite{Fiedlschuster_PhD2018, Schmidt_PRA2017} in various situations where nonadiabatic effects are induced by an external time-dependent field, either laser pulse or continuous wave laser, that such TDPES very much resembles Floquet surfaces, rather than Born-Oppenheimer surfaces or quasi-static surfaces~\cite{Vrakking_PCCP2011, Suzuki_PCCP2015}. Therefore, it seems natural to employ the Floquet representation of electronic dynamics driven by an external time-periodic field in combination with CT-MQC. Note that, despite the fact that CT-MQC has been derived from the exact-factorization electronic and nuclear equations, electronic dynamics is solved in a basis. The choice of the ``most suitable'' electronic basis is, thus, crucial. It is worth mentioning here that other trajectory-based approaches to excited-state dynamics have been combined with the Floquet formalism in order to explicitly include the effect of the photo-exciting or driving field, as the quantum-classical Liouville equation~\cite{Schuette_JCP2001}, ab initio multiple spawning~\cite{Bucksbaum_JPB2015}, ab initio multiple cloning~\cite{Shalashilin_CP2018}, and trajectory surface hopping~\cite{Subotnik_JCP2020, Subotnik_JCTC2020, Gonzalez_JPCA2012, Schmidt_PRA2016}. 

The present work develops and applies CT-MQC using the Floquet formalism (F-CT-MQC). To this end, in Section~\ref{sec: theory}, we briefly recall the exact factorization, introduce the elements of the Floquet theory necessary for F-CT-MQC, and derive F-CT-MQC equations. In Section~\ref{sec: results}, F-CT-MQC is applied to simulate the periodically-driven, nonadiabatic dynamics of a model system, for which the exact solution is available, allowing us to test the performance of the algorithm. Our conclusions are presented in Section~\ref{sec: conclusions}.

\section{Exact factorization for periodically driven systems: The Floquet picture}\label{sec: theory}

We study a system of interacting electrons and nuclei subject to an external time-dependent classical field $\hat{V}(\mathbf r,\mathbf R,t)$. The Hamiltonian describing the system is
\begin{align}\label{eqn: H}
\hat H(\mathbf r,\mathbf R,t) = \hat T_n(\mathbf R)+\hat H_{el}(\mathbf r,\mathbf R)+\hat{V}(\mathbf r,\mathbf R,t)
\end{align}
with electronic positions labelled by $\mathbf r$ and nuclear positions labelled by $\mathbf R$. The field-free molecular Hamiltonian is the sum of the nuclear kinetic energy $\hat T_n(\mathbf R)$ and of the electronic Hamiltonian $\hat H_{el}(\mathbf r,\mathbf R)$ which is the sum of the electronic kinetic energy and the interaction potential. 

The solution of the time-dependent Schr\"{o}dinger equation (tdSE), $\Psi(\mathbf r,\mathbf{R},t)$, with Hamiltonian~(\ref{eqn: H}) can be factored as the product of a nuclear wavefunction, $\chi(\mathbf R,t)$, and an electronic conditional factor, $\Phi(\mathbf r,t;\mathbf R)$, that parametrically depends on $\mathbf R$, namely~\cite{Gross_PRL2010, EF_bookchapter_2020}
\begin{align}\label{eqn: EF}
\Psi(\mathbf r,\mathbf{R},t) = \chi(\mathbf R,t)\Phi(\mathbf r,t;\mathbf R)
\end{align}
The evolution equations for $\chi(\mathbf R,t)$ and $\Phi(\mathbf r,t;\mathbf R)$ are derived from the full  tdSE~\cite{Gross_JCP2012, Alonso_JCP2013, Gross_JCP2013, Ciccotti_EPJB2018} and are
\begin{align}
i\hbar \partial_t \chi(\mathbf R,t) &= \left[\frac{1}{2}\mathbf M^{-1}\left[-i\hbar\boldsymbol{\nabla} + \mathbf A(\mathbf R,t)\right]^2 + \epsilon(\mathbf R,t)+ \epsilon_{\mathrm{ext}}(\mathbf R,t)\right] \chi(\mathbf R,t)\label{eqn: EF n} \\
i\hbar \partial_t  \Phi(\mathbf r,t;\mathbf R) &= \left[\hat H_{el}(\mathbf r,\mathbf R)+\hat{V}(\mathbf r,\mathbf R,t)+\hat U\left[\Phi,\chi\right]-\epsilon(\mathbf R,t)-\epsilon_{\mathrm{ext}}(\mathbf R,t)\right]\Phi(\mathbf r,t;\mathbf R)\label{eqn: EF el}
\end{align}
where the symbol $\mathbf M$ stands for the diagonal (constant) mass tensor and $\boldsymbol\nabla$ for the spatial derivative with respect to nuclear positions. The time-dependent potentials of the exact factorization mediate the coupling between electrons and nuclei, and are the time-dependent vector potential~\cite{Requist_PRA2015, Requist_PRA2017, Agostini_JPCL2017, Curchod_EPJB2018}
\begin{align}\label{eqn: TDVP}
\mathbf A(\mathbf R,t) = \left\langle\Phi(\mathbf r,t;\mathbf R)\right| \left.-i\hbar\boldsymbol{\nabla}\Phi(\mathbf r,t;\mathbf R)\right\rangle
\end{align}
and the time-dependent scalar potential, or time-dependent potential energy surface (TDPES)~\cite{Gross_PRL2013, Gross_MP2013, Min_PRL2014, Gross_JCP2015, Curchod_JCP2016, Maitra_PRL2019}, which we decompose into two contributions, namely
\begin{align}\label{eqn: TDPES}
\epsilon(\mathbf R,t) = \left\langle\Phi(\mathbf r,t;\mathbf R)\right|\hat H_{el}(\mathbf r,\mathbf R)+\hat U\left[\Phi,\chi\right]-i\hbar \partial_t\left|\Phi(\mathbf r,t;\mathbf R)\right\rangle
\end{align}
and
\begin{align}\label{eqn: TDPES ext}
\epsilon_{\mathrm{ext}}(\mathbf R,t) = \left\langle\Phi(\mathbf r,t;\mathbf R)\right|\hat{V}(\mathbf r,\mathbf R,t)\left|\Phi(\mathbf r,t;\mathbf R)\right\rangle
\end{align}
The electron-nuclear coupling operator $\hat U\left[\Phi,\chi\right]=\hat U\left[\Phi(\mathbf r,t;\mathbf R),\chi(\mathbf R,t)\right]$ is~\cite{Gross_EPL2014, Gross_JCP2014, Agostini_ADP2015}
\begin{align}\label{eqn: enco}
\hat U [\Phi,\chi]=& \mathbf M^{-1}\Bigg[\frac{1}{2}[-i\hbar \boldsymbol{\nabla}+\mathbf A(\mathbf R,t)]^2+\left(\frac{-i\hbar\boldsymbol{\nabla}\chi(\mathbf R,t)}{\chi(\mathbf R,t)}+\mathbf A(\mathbf R,t)\right)\cdot\big(-i\hbar\boldsymbol{\nabla}-\mathbf A(\mathbf R,t)\big)\Bigg]
\end{align}
and depends explicitly on the nuclear wavfunction $\chi(\mathbf R,t)$, and implicitly on the electronic factor $\Phi(\mathbf r,t;\mathbf R)$, via its dependence on the vector potential. The integration operation, indicated as $\langle \,\cdot \,\rangle$ in previous equations, will be discussed below.  

While the derivation just presented is valid in general situations, we focus here on the case of an external drive $\hat V(\mathbf r,\mathbf R,t)$, with constant amplitude and periodic in time with frequency $\Omega=2\pi/T$ (continuous wave (cw) laser). In such case, we will generalize the trajectory-based approach developed to solve Eqs.~(\ref{eqn: EF n}) and~(\ref{eqn: EF el}), and dubbed coupled-trajectory mixed quantum-classical (CT-MQC) algorithm, to the Floquet formalism.

\subsection{Elements of Floquet theory for CT-MQC}
The implementation of CT-MQC relies on the expansion of the electronic wavefunction $\Phi(\mathbf r,t;\mathbf R)$ on an electronic basis. Since the system is subject to a time-periodic Hamiltonian it seems natural to expand on a basis that takes already into account, to some extent, the driven nature of the problem. As we discuss below this can be done using Floquet theory and will naturally lead to two different classes of states  that can be used as electronic basis, the Floquet adiabatic and diabatic (or dressed) states, that we briefly introduce here.

The electronic time-dependent periodic Hamiltonian is 
$\hat H_{el}(\mathbf r,\mathbf R)+\hat{V}(\mathbf r,\mathbf R,t)$, at fixed nuclear positions.  As such, according to the Floquet theorem~\cite{Floquet_1880,Sambe_PRA1973,Hanggi_PhysRep98}, a complete set of solutions of the electronic tdSE, if there were no nonadiabatic coupling to the nuclei, takes the form $e^{i\mathcal E_{\alpha}(\mathbf R)t}\phi_{\alpha}(\mathbf r,t;\mathbf R)$, where $\phi_{\alpha}(\mathbf r,t;\mathbf R)$ are the so called Floquet \emph{adiabatic states} which are constructed as the eigenstates of the Floquet Hamiltonian 
\begin{align}
\hat H_{Fl}(\mathbf r, \mathbf R,t)=\hat H_{el}(\mathbf r,\mathbf R)+\hat{V}(\mathbf r,\mathbf R,t)- i\hbar\partial_t
\end{align} 
i.e., they satisfy the eigenvalue problem
\begin{align}\label{eqn: Fl adiabatic eqn}
\hat H_{Fl}(\mathbf r, \mathbf R,t) \phi_{\alpha}(\mathbf r,t;\mathbf R)  = \mathcal E_{\alpha}(\mathbf R) 
\phi_{\alpha}(\mathbf r,t;\mathbf R) 
\end{align}
In the equation above the eigenvalues $\mathcal E_{\alpha}(\mathbf R)$ are called Floquet quasi-energies and do not depend on time (but they depend on $\mathbf R$ as effect of the parametric dependence of the electronic Hamiltonian on the nuclear positions); $\phi_{\alpha}(\mathbf r,t;\mathbf R) = \phi_{\alpha}(\mathbf r,t+T;\mathbf R)$ has the same periodicity of the external drive. Given the periodicity of the Floquet states $\phi_{\alpha}(\mathbf r,t;\mathbf R)$, we can expand them in harmonics of the drive, as
\begin{align}
\phi_{\alpha}(\mathbf r,t;\mathbf R) = \sum_{n=-\infty}^{n=+\infty} e^{i\omega_nt}\phi_{\alpha,n}(\mathbf r;\mathbf R)
\end{align}
with $\omega_n=n\Omega$ and $n$ an integer.  Inserting this expression into Floquet equation~(\ref{eqn: Fl adiabatic eqn}) and projecting onto the harmonic $m$ we get
\begin{align}\label{eqn: ad_floquet_harmonics}
\sum_n \left[\left(\hat H_{el}(\mathbf r,\mathbf R) +\hbar\omega_m\hat I\right)\delta_{mn} + \hat V_{mn}(\mathbf r,\mathbf R)\right] \phi_{\alpha,n}(\mathbf r;\mathbf R) = \mathcal E_{\alpha}(\mathbf R)\phi_{\alpha,m}(\mathbf r;\mathbf R)
\end{align}
where 
\begin{align}
\hat{V}_{mn}(\mathbf r,\mathbf R)=\frac{1}{T}\int_0^T\,dt \;e^{i\left(\omega_n-\omega_m\right)t}\,\hat{V}(\mathbf r,\mathbf R,t)\,.
\end{align}
We recognize in Eq.~(\ref{eqn: ad_floquet_harmonics}) an eigenvalue problem in an extended space, including both the electronic degrees of freedoms as well as the harmonics of the drive. Note that, here, the scalar product in the space of the harmonics is defined via a time integral over a period of the drive and satisfies the orthonormality condition 
$$
\frac{1}{T}\int_0^T\,dt \;e^{i\left(\omega_n-\omega_m\right)t}=\delta_{nm}
$$
The Floquet adiabatic states obtained solving the eigenproblem~(\ref{eqn: ad_floquet_harmonics}) have a mixed electronic-field nature since $\hat{V}_{mn}(\mathbf r,\mathbf R)$ couples different harmonics, while the operator $\hat H_{el}(\mathbf r,\mathbf R) +\hbar\omega_m\hat I$ is diagonal in the space of harmonics.
We can now expand the electronic wavefunction of the exact factorization in the basis of Floquet adiabatic states as
\begin{align}\label{eqn:expansion_adiafloq}
\Phi(\mathbf r,t;\mathbf R)=\sum_\alpha C_\alpha(\mathbf R,t) \phi_{\alpha}(\mathbf r,t;\mathbf R)=
\sum_{\alpha }\sum_n C_{\alpha}\big(\mathbf R,t\big)e^{i\omega_nt}\phi_{\alpha,n}(\mathbf r;\mathbf R)
\end{align}
where the sum over $\alpha$ runs over the complete set of Floquet states, solution of Eq.~(\ref{eqn: Fl adiabatic eqn}), and the expansion coefficients $C_{\alpha}(\mathbf R,t)$ depend on the nuclear positions as well as on time. Note that, while the Floquet adiabatic states $\phi_{\alpha}(\mathbf r,t;\mathbf R)$ have periodicity $T$, this is in general not the case for the electronic wavefunction  $\Phi(\mathbf r,t;\mathbf R)$, which is obtained from the full electron-nuclear wavefunction of the problem (for which one could also use Floquet theorem) through the exact factorization ansatz in Eq.~(\ref{eqn: EF}). This observation justifies the use of time-dependent coefficients in the expansion in Eq.~(\ref{eqn:expansion_adiafloq}).

As we have seen so far, constructing the Floquet adiabatic states requires the solution of an eigenvalue problem in an extended space, which can be a burden for realistic applications beyond model systems.  A different set of states, called Floquet \emph{diabatic states}, can be obtained considering the electronic field-free Floquet Hamiltonian
\begin{align}\label{eqn: FH diabatic eqn}
\hat H_{Fl}^{(0)}(\mathbf r,\mathbf R,t) = \hat H_{el}(\mathbf r, \mathbf R) - i\hbar\partial_t
\end{align}
Following from Floquet theorem~\cite{Sambe_PRA1973}, the eigenvalue equation
\begin{align}\label{eqn: Fl diabatic eqn}
\hat H_{Fl}^{(0)}(\mathbf r, \mathbf R,t) \varphi_{\alpha}(\mathbf r,t;\mathbf R)  = \mathcal E_{\alpha}^{(0)}(\mathbf R) \varphi_{\alpha}(\mathbf r,t;\mathbf R) 
\end{align}
yields the Floquet eigenmodes $\varphi_{\alpha}(\mathbf r,t; \mathbf R)$ and the Floquet quasi-energies $\mathcal E_{\alpha}^{(0)}(\mathbf R)$. As for the adiabatic modes, the diabatic eigenmodes are periodic in time $\varphi_{\alpha}(\mathbf r,t+T;\mathbf R)=\varphi_{\alpha}(\mathbf r,t;\mathbf R)$ with the same period $T$ of the drive. However, as compared to the Floquet states introduced previously, we expect the diabatic states to have a much simpler structure in the space of harmonics, since by construction the time-dependent drive is not included in the Floquet Hamiltonian of Eq.~(\ref{eqn: FH diabatic eqn}). In fact, we can write explicitly the expression of these Floquet diabatic states, in terms of the standard Born-Oppenheimer basis $\lbrace\psi_{k}(\mathbf r;\mathbf R)\rbrace_{k=1,\ldots,+\infty}$, defined as the eigenstates of the electronic Hamiltonian $\hat H_{el}(\mathbf r,\mathbf R)$ for each nuclear configuration $\mathbf R$ and satisfying the condition $\hat H_{el}(\mathbf r,\mathbf R)\psi_{k}(\mathbf r;\mathbf R) = E_k(\mathbf R)\psi_{k}(\mathbf r;\mathbf R)$. A direct calculation shows that the state
 \begin{align}
 \varphi_{\alpha}(\mathbf r,t; \mathbf R)  = e^{i\omega_n t}\psi_{k}(\mathbf r;\mathbf R),\quad \alpha=(k,n)
\end{align}  
is eigenstate of the drive-free Floquet Hamiltonian~(\ref{eqn: FH diabatic eqn}) with quasi-energy
 $\mathcal E_{\alpha}^{(0)}(\mathbf R)=E_k(\mathbf R)+\hbar\omega_n$. The Floquet diabatic potential energy surfaces (PESs) are simply the standard Born-Oppenheimer PESs rigidly shifted by the energy of the corresponding harmonic.

Similarly to Eq.~(\ref{eqn:expansion_adiafloq}), we can choose the Floquet diabatic states as electronic representation for the electronic wavefunction $\Phi(\mathbf r,t;\mathbf R)$, and write
\begin{align}\label{eqn: Fl dia Phi}
\Phi(\mathbf r,t;\mathbf R) = \sum_k \sum_n C_{k,n}\big(\mathbf R,t\big)e^{i\omega_nt}\psi_{k}\big(\mathbf r;\mathbf R\big)
\end{align}
where the two sums run over the electronic physical states $k$ and over the harmonics $n$. As observed previously, we allow an explicit time-dependence for the coefficients $C_{k,n}\big(\mathbf R,t\big)$ in order to reproduce the  dynamics of the electronic wavefunction which does not need to be periodic in time with the period of the drive.

While a priori the Floquet adiabatic and diabatic bases could be used in trajectory-based calculations, and one might expect the Floquet adiabatic basis to perform better since it contains already information on the external drive, we will argue that for the practical implementation of F-CT-MQC, the Floquet diabatic basis is preferred. For this reason in the next section we will derive F-CT-MQC equations only for this basis, although the derivation for the Floquet adiabatic basis can be obtained via an obvious generalization.

\subsection{F-CT-MQC with Floquet diabatic states}\label{sec: F-CT-MQC}
We introduce the trajectory-based representation of nuclear dynamics~\cite{Agostini_JCTC2020_1} that follows from the exact-factorization equations~(\ref{eqn: EF n}) and~(\ref{eqn: EF el}). The nuclear wavefunction is written in polar representation as $\chi(\mathbf R,t)= |\chi(\mathbf R,t)|\exp{[(i/\hbar)S(\mathbf R,t)]}$, such that the coupled evolution equations
\begin{align}
-\partial_t S(\mathbf R,t) &=\mathbf M^{-1}\frac{\left[\boldsymbol{\nabla} S(\mathbf R,t)+\mathbf A(\mathbf R,t)\right]^2}{2}+\epsilon(\mathbf R,t)+\epsilon_{\mathrm{ext}}(\mathbf R,t)\label{eqn: HJ}\\
\partial_t\left|\chi(\mathbf R,t)\right|^2&= -\boldsymbol{\nabla} \cdot \left[\mathbf M^{-1}\left(\boldsymbol{\nabla}  S(\mathbf R,t)+\mathbf A(\mathbf R,t)\right)\left|\chi(\mathbf R,t)\right|^2\right]\label{eqn: continuity}
\end{align}
are derived from the nuclear time-dependent Schr\"odinger equation~(\ref{eqn: EF n}). The equation for the phase $S(\mathbf R,t)$ is given in the classical limit by neglecting in Eq.~(\ref{eqn: HJ}) the quantum potential term. The first equation can be solved with the method of characteristics, introducing a set of ordinary differential equations -- the characteristic equations --  yielding the values of the field $S(\mathbf R,t)$ $\forall\,\, \mathbf R,t$. The characteristic equations are Hamilton-like equations that describe the evolution in time of the ``variables'' $\mathbf R(t)$ and $\mathbf P(t)\equiv\boldsymbol{\nabla} S(\mathbf R(t),t)+\mathbf A(\mathbf R(t),t)$ appearing in Eq.~(\ref{eqn: HJ}), i.e.,
\begin{align}
\dot{\mathbf R}(t) &=\mathbf M^{-1} \mathbf P(t) \label{eqn: R dot}\\
\dot{\mathbf P}(t) &=-\boldsymbol{\nabla} \left(\epsilon\big(\mathbf R(t),t\big) +\epsilon_{\mathrm{ext}}\big(\mathbf R(t),t\big)+ \left[\dot{\mathbf R}(t)\cdot\mathbf A\big(\mathbf R(t),t\big)\right]\right)+ \dot{\mathbf A}\big(\mathbf R(t),t\big)\label{eqn: P dot}
\end{align}
The procedure to derive Eqs.~(\ref{eqn: R dot}) and~(\ref{eqn: P dot}) has been illustrated in detail in Refs.~\cite{Ciccotti_EPJB2018, Ciccotti_JPCA2020}, therefore, we just mention here that the first term on the right-hand side of Eq.~(\ref{eqn: P dot}) is the gradient of the (pseudo-classical) Hamiltonian defined on the right-hand side of Eq.~(\ref{eqn: HJ}). The time-dependent potentials $\epsilon\big(\mathbf R(t),t\big)$, $\epsilon_{\mathrm{ext}}\big(\mathbf R(t),t\big)$ and $\mathbf A\big(\mathbf R(t),t\big)$ are evaluated along each characteristic $\mathbf R(t)$ by solving the electronic equation~(\ref{eqn: EF el}) along the same characteristic. Solving the partial differential equation~(\ref{eqn: EF el}) along the \textsl{flow} of trajectories -- the characteristics -- requires to switch from the Eulerian frame to the Lagrangian frame, but before discussing the electronic equation, let us discuss the continuity equation~(\ref{eqn: continuity}).

The evolution of the nuclear density, Eq.~(\ref{eqn: continuity}), is described by a ``standard'' continuity equation, that can be solved coupled to Eqs.~(\ref{eqn: R dot}) and~(\ref{eqn: P dot}). This would allow us to reconstruct the nuclear wavefunction, the only approximation being that we dropped the quantum potential in Eq.~(\ref{eqn: HJ}). Neglecting the quantum potential has the effect of decoupling the evolution of the phase from the evolution of the density, while the continuity equation still depends on the phase. Therefore, as done in previous work~\cite{Ciccotti_EPJB2018, Gross_JCTC2016}, we will not solve the continuity equation, and we will only reconstruct a classical-like nuclear density from the distribution of the trajectories, working in the hypothesis that for short enough times, an ensemble of trajectories evolving with Eqs.~(\ref{eqn: R dot}) and~(\ref{eqn: P dot}) will sample portions of nuclear configuration space with high probability density.

Equations~(\ref{eqn: R dot}) and~(\ref{eqn: P dot}) are the basis of the CT-MQC algorithm, that has been derived and tested in Refs.~\cite{Gross_PRL2015, Gross_JCTC2016}, applied to study the relaxation dynamics through conical intersections in Refs.~\cite{Agostini_JCTC2020_2, Agostini_JCP2021}, and in Refs.~\cite{Gross_JPCL2017, Tavernelli_EPJB2018} in combination with time-dependent density functional theory to study the photo-induced dynamics in oxirane, and thoroughly analyzed in Refs.~\cite{Agostini_EPJB2018, Maitra_JCTC2018}.

The exact-factorization product form of the solution of the tdSE is invariant under a $(\mathbf R,t)$-dependent phase transformation of the nuclear and electronic wavefunctions. Therefore, the gauge freedom has to be fixed by choosing the gauge. As previously done, in CT-MQC the gauge is chosen such that $\epsilon\big(\mathbf R(t),t\big)+\dot{\mathbf R}(t)\cdot\mathbf A\big(\mathbf R(t),t\big)=0$. With this choice of gauge, the characteristic equation~(\ref{eqn: P dot}) for the nuclear momentum yields
\begin{align}\label{eqn: classical force 1}
\dot{\mathbf P}(t) &=-\boldsymbol{\nabla} \epsilon_{\mathrm{ext}}\big(\mathbf R(t),t\big)+ \dot{\mathbf A}\big(\mathbf R(t),t\big)
\end{align}
which gives the expression of the classical force $\dot{\mathbf P}(t) = \mathbf F(t)$ that drives F-CT-MQC trajectories. The explicit form of each term will be given below, after determining the evolution equation for the electronic wavefunction.

The electronic equation~(\ref{eqn: EF el}) is expressed along the characteristics $\mathbf R(t)$  as
\begin{align}\label{eqn: dot Phi}
i\hbar\dot{\Phi}(\mathbf r,t;\mathbf R(t)) = \left[\hat H_{el}\big(\mathbf r,\mathbf R(t)\big)+\hat V\big(\mathbf r,\mathbf R(t),t\big)+\hat U\left[\Phi,\chi\right]-\epsilon\big(\mathbf R(t),t\big)-\epsilon_{\mathrm{ext}}\big(\mathbf R(t),t\big)\right.\nonumber \\
\left.+i\hbar\dot{\mathbf R}(t)\cdot\boldsymbol{\nabla} \right]\Phi(\mathbf r,t;\mathbf R(t))
\end{align}
Switching from the Eulerian to the Lagrangian frame, only total time derivatives can be evaluated \textsl{along the flow}, that is why the symbol $\dot{\Phi}(\mathbf r,t;\mathbf R(t))$ has been introduced. Furthermore, we used the relation $\dot\Phi(\mathbf r,t;\mathbf R(t)) = \partial_t\Phi(\mathbf r,t;\mathbf R(t))+ \dot{\mathbf R}(t)\cdot \boldsymbol{\nabla}\Phi(\mathbf r,t;\mathbf R(t))$ to replace the partial time derivative of Eq.~(\ref{eqn: EF el}) with the total time derivative, and thus obtain the last term in square brackets on the right-hand side of Eq.~(\ref{eqn: dot Phi}). The expansion of the electronic wavefunction in the Floquet diabatic basis is expressed as well along a trajectory $\mathbf R(t)$
\begin{align}\label{eqn: Fl dia Phi with R(t)}
\Phi(\mathbf r,t;\mathbf R(t)) = \sum_k \sum_n C_{k,n}\big(\mathbf R(t),t\big)e^{i\omega_nt}\psi_{k}\big(\mathbf r;\mathbf R(t)\big)
\end{align}
and is inserted into Eq.~(\ref{eqn: dot Phi}). When using it in the left-hand side of Eq.~(\ref{eqn: dot Phi}), we have
\begin{align}
\dot{\Phi}(\mathbf r,t;\mathbf R(t))= \sum_{k,n} \left[\dot C_{k,n}\big(\mathbf R(t),t\big)\psi_k(\mathbf r;\mathbf R(t)) + C_{k,n}\big(\mathbf R(t),t\big) \dot\psi_k(\mathbf r;\mathbf R(t))\right.\nonumber \\
\left.+i\omega_nC_{k,n}\big(\mathbf R(t),t\big) \psi_k(\mathbf r;\mathbf R(t))\right]e^{i\omega_nt}
\end{align}
Applying the total time derivative to the electronic state in the second term on the right-hand side, we find that the partial time derivative is zero because $\psi_k(\mathbf r;\mathbf R(t))$ depends on time only via its dependence on the trajectory, then
\begin{align}\label{eqn: Phi dot with Floquet diabatic}
\dot{\Phi}(\mathbf r,t;\mathbf R(t))=& \sum_{k,n} \left[\dot C_{k,n}\big(\mathbf R(t),t\big) 
+C_{k,n}\big(\mathbf R(t),t\big) \dot{\mathbf R}(t)\cdot \boldsymbol{\nabla}+i\omega_nC_{k,n}\big(\mathbf R(t),t\big) \right]\psi_k(\mathbf r;\mathbf R(t)) e^{i\omega_nt}
\end{align}
With the aim to derive the evolution equations for $C_{k,n}\big(\mathbf R(t),t\big)$, the expansion in Floquet diabatic states is used as well in the right-hand side of Eq.~(\ref{eqn: dot Phi}). The whole procedure is detailed in  Appendix~\ref{app: el eqn}, and we discuss here only the main differences if compared to the standard derivation of CT-MQC equations.

In order to isolate in Eq.~(\ref{eqn: dot Phi}) a term $\dot C_{l,m}\big(\mathbf R(t),t\big) $, it is necessary to project onto the state $e^{i\omega_mt}\psi_{l}\big(\mathbf r;\mathbf R(t)\big)$. Working in the Floquet basis, the projection operation involves an integral over time, as well as an integral over the electronic configuration space. To do this, we introduce the change of variable $t\rightarrow s$ in $e^{i\omega_nt}$ (and $e^{i\omega_mt}$). This is done to separate the time dependence in the electronic equation into a periodic contribution induced by the drive, and indicated with $s$, and a non-periodic contribution, indicated with $t$. The time $s$ is only used when the integral over the period is computed, in order to distinguish it from the non-periodic time dependence $t$. After the projection operation, the dependence on $s$ disappears because it has been integrated out over a period, but the dependence on $t$ remains. The projection operation onto the state $e^{i\omega_ms}\psi_{l}\big(\mathbf r;\mathbf R(t)\big)$ is performed in the extended harmonic-electronic space, with the time integration over a period and the spatial integration over the whole electronic configuration space. Therefore, the instantaneous expectation value on the electronic wavefunction $\Phi(\mathbf r,t;\mathbf R(t))$  of a general electron-nuclear operator $\hat O(\mathbf r,\mathbf R(t))$ that does not depend explicitly on time is computed as
\begin{align}
&\left\langle \Phi(\mathbf r,t;\mathbf R(t))\left|\hat O\big(\mathbf r,\mathbf R(t)\big)\right|\Phi(\mathbf r,t;\mathbf R(t)) \right\rangle =\nonumber\\
& \frac{1}{T}\int_0^T ds \int d\mathbf r \sum_{k,n}\sum_{l,m}\bar C_{k,n}\big(\mathbf R(t),t\big) e^{-i\omega_n s} \bar\psi_k(\mathbf r;\mathbf R(t)) \hat O\big(\mathbf r,\mathbf R(t)\big)C_{l,m}\big(\mathbf R(t),t\big) e^{i\omega_m s} \psi_l(\mathbf r;\mathbf R(t))\label{eqn: average over s and r}
\end{align}
showing that the periodic time dependence carried by the harmonics and the variable $s$ is integrated out, whereas the time dependence in $t$ remains. This operation is applied in the definition of the TDPES and of the time-dependent vector potential. Note that, this separation of times is crucial to keep working in a time-dependent picture while using the Floquet formalism. As discussed in the Introduction, the Floquet picture maps a periodic time-dependent problem into an effective time-independent problem in an enlarged harmonic-electronic vector space, thus reducing the solution of the time evolution to a diagonalization in this extended space. When separating the -- full -- molecular time-periodic process into a coupled electron-nuclear process to apply trajectory-based quantum-classical schemes, the time periodicity is only partially accounted for in the definition of the Floquet diabatic states, because, in general, the electronic and the nuclear dynamics are not time periodic (only the full molecular wavefunction is). It should be mentioned that the benefit of this procedure has been identified in previous work on the combination of the Floquet formalism with the quantum-classical Liouville equation~\cite{Schuette_JCP2001}. The identification of two time scales allows one to account for the oscillatory time dependence induced by the drive, while the remaining part of the time evolution is explicitly accounted for via the evolution of the expansion coefficients (as in Eq.~(\ref{eqn: Fl dia Phi with R(t)})). Essentially, the operation described in Eq.~(\ref{eqn: average over s and r}) allows us to obtain time-dependent quantities after their dependence on the driving frequency of the laser has been averaged out, when considering their effect on the nuclei; the effect of the driving frequency on the electrons is treated in Fourier space, as it is treated in terms of Floquet diabatic states and transitions among them. A similar separation of time scales has been used in Floquet surface hopping~\cite{Schmidt_PRA2016} with the aim to extend the Floquet picture to processes that are not strictly periodic, as for instance nonadiabatic processes initiated by a laser pulse.

Henceforth, the dependence on the trajectories $\mathbf R(t)$ will be replaced by a superscript $\nu$, which stands for a trajectory index -- this is done mainly to simplify the notation used in the equations below. The solution of Eq.~(\ref{eqn: HJ}) with the method of characteristics requires to solve Eqs.~(\ref{eqn: R dot}) and~(\ref{eqn: P dot}) for a large number of initial conditions, $\nu=1,\ldots,N_{tr}$. The index $\nu$ is, thus, used to label the trajectories, or characteristics. Similarly to the original form of the electronic evolution equation of CT-MQC~\cite{Maitra_JCTC2018, Gross_JCTC2016, Agostini_EPJB2018, Agostini_JCTC2020_1}, the final expression of the time derivative of the Floquet diabatic coefficients, derived in Appendix~\ref{app: el eqn}, is
\begin{align}\label{eqn: dot Csd 3}
\dot C_{l,m}^\nu(t) = \dot C_{l,m}^\nu(t)\Big|_{\mathrm{EH}} + \dot C_{l,m}^\nu(t)\Big|_{\mathrm{QM}} + \dot C_{l,m}^\nu(t)\Big|_{\mathrm{EXT}}
\end{align}
which has to be solved along a trajectory $\nu$. The three contributions identified in Eq.~(\ref{eqn: dot Csd 3}) are: an Ehrenfest-like term (EH)
\begin{align}\label{eqn: dot Csd 3a}
\dot C_{l,m}^\nu(t)\Big|_{\mathrm{EH}}=-\frac{i}{\hbar}\left(E_l^\nu+\hbar\omega_m\right)C_{l,m}^\nu(t)-\dot{\mathbf R}^\nu(t) \cdot\sum_{k,n} \mathbf d_{lk}^\nu\,C_{k,n}^\nu(t)\delta_{mn}
\end{align}
a term depending on the quantum momentum (QM)
\begin{align}\label{eqn: dot Csd 3b}
\dot C_{l,m}^\nu(t)\Big|_{\mathrm{QM}}= \frac{1}{\hbar}\left[\mathbf M^{-1}\boldsymbol{\mathcal P}^\nu(t)\right]\cdot\left[\mathbf f_{l,m}^\nu-\mathbf A^\nu(t)\right]C_{l,m}^\nu(t)
\end{align}
and a term with explicit dependence on the external field
\begin{align}\label{eqn: dot Csd 3c}
\dot C_{l,m}^\nu(t)\Big|_{\mathrm{EXT}}=-\frac{i}{\hbar}\epsilon_{\mathrm{ext}}^\nu(t)C_{l,m}^\nu(t) -\sum_{k,n}\frac{i}{\hbar} V_{mn,lk}^\nu C_{k,n}^\nu(t)
\end{align}
Note that, the only difference between the expression given in~(\ref{eqn: dot Csd 3}) and the field-free CT-MQC electronic equation resides in the additional terms depending on the external field (EXT), via $\epsilon_{\mathrm{ext}}^\nu(t)$ and $V_{mn,lk}^\nu$ given in Eq.~(\ref{eqn: dot Csd 3c}). The contribution $\epsilon_{\mathrm{ext}}^\nu(t)$ to the TDPES in the Floquet diabatic basis reads
\begin{align}\label{eqn: ext TDPES with states}
\epsilon_{\mathrm{ext}}^\nu(t) = \sum_{k,l} \sum_{n,m} \bar C_{k,n}^\nu(t) C_{l,m}^\nu(t)V_{mn,lk}^\nu
\end{align}
where the matrix elements of the driving field are
\begin{align}\label{eqn: ext field matrix}
V_{mn,lk}^\nu=\frac{1}{T}\int_0^T\,ds \int d\mathbf r\;e^{i\left(\omega_n-\omega_m\right)s}\,\bar{\psi}_l^\nu(\mathbf r)\hat{V}^\nu(\mathbf r,s)\psi_k^\nu(\mathbf r)
\end{align}
As done above, all quantities depending on $\mathbf R(t)$ have acquired a superscript $\nu$. The ``periodic'' time dependence has been indicated with the symbol $s$, which is integrated over a period in the projection operation. Combining Eqs.~(\ref{eqn: ext TDPES with states}) and~(\ref{eqn: ext field matrix}) shows that the expression of $\epsilon_{\mathrm{ext}}^\nu(t)$ easily follows from its definition in Eq.~(\ref{eqn: TDPES ext}). Since only the periodic time dependence is integrated out, $\epsilon_{\mathrm{ext}}^\nu(t)$ still depends on time $t$. The quantity indicated as $V_{mn,lk}^\nu$  stands for the Fourier transform of the matrix elements of the external drive, from Eq.~(\ref{eqn: ext field matrix}). The term $V_{mn,lk}^\nu$ induces population transfer along the trajectory $\nu$ between different electronic states and different harmonics as effect of the action of the external field. In particular, if only one driving frequency is considered, as in the present case, the non-zero matrix elements of $V_{mn,lk}^\nu$ are those satisfying the condition $m-n=\pm1$.

In Eqs.~(\ref{eqn: dot Csd 3a}) and~(\ref{eqn: dot Csd 3b}), the ``standard'' terms of the electronic evolution equation of F-CT-MQC contain: the time-dependent vector potential $\mathbf A^\nu(t)$, the Floquet diabatic force $\mathbf f_{l,m}^\nu$ accumulated along the trajectory $\nu$, the quantum momentum $\boldsymbol{\mathcal P}^\nu(t)$, the classical velocity of the trajectory $\dot{\mathbf R}^\nu(t)$, and the nonadiabatic coupling vectors $ \mathbf d_{lk}^\nu$. In Eq.~(\ref{eqn: dot Csd 3b}), the time-dependent vector potential is
\begin{align}
\mathbf A^\nu(t) =  \frac{1}{T}\int_0^Tds\int d\mathbf r\sum_{k,l} \sum_{n,m} \bar C_{k,n}^\nu(t) e^{-i\omega_ns}\bar{\psi}_{k}^\nu(\mathbf r)\left(-i\hbar\boldsymbol{\nabla}\right)C_{l,m}^\nu(t)e^{i\omega_ms}\psi_{l}^\nu(\mathbf r)\label{eqn: tdvp in basis}
\end{align}
following its definition in Eq.~(\ref{eqn: TDVP}). Also in this case, the average operation is performed over the periodic time dependence, such that $\mathbf A^\nu(t)$ still depends on time $t$. The quantity indicated with the symbol $\mathbf f_{l,m}^\nu$ in Eq.~(\ref{eqn: dot Csd 3b}) is the force from the Floquet diabatic PES $E_l^\nu+\hbar\omega_m$ accumulated over time up to time $t$ along the trajectory $\nu$. Since the Floquet diabatic PESs are parallel to the adiabatic PESs, such an accumulated force can be computed from the gradient of $E_l^\nu$ as in standard CT-MQC. Such an accumulated force is used in the F-CT-MQC-approximate expression of the time-dependent vector potential, namely
\begin{align}
\mathbf A^\nu(t) = \sum_{l,m} \left|C_{l,m}^\nu(t)\right|^2\mathbf f_{l,m}^\nu
\end{align}
which is related to the gradient of the harmonic-electronic coefficients $C_{l,m}^\nu(t)$ of Eq.~(\ref{eqn: tdvp in basis}) (see for details Refs.~\cite{Gross_PRL2016, Gross_JCTC2016, Agostini_JCTC2020_1}). In Eqs.~(\ref{eqn: dot Csd 3a}) and~(\ref{eqn: dot Csd 3b}), the nuclear velocity $\dot{\mathbf R}^\nu(t)$ and the quantum momentum $\boldsymbol{\mathcal P}^\nu(t)$ appear, via the contribution of the electron-nuclear coupling operator~(\ref{eqn: enco}) that depends on the nuclear wavefunction (as discussed in Appendix~\ref{app: el eqn}). The nuclear velocity in Eq.~(\ref{eqn: dot Csd 3a}) couples to the nonadiabatic coupling vectors $\mathbf d_{lk}^\nu$, also known as derivative couplings, driving the population transfer between the electronic states $k$ and $l$ that belong to the same harmonic ($\delta_{mn}$), as effect of the nuclear displacement operator $\boldsymbol{\nabla}$. They can be evaluated numerically or analytically via the Hellmann-Feynman formula, exactly as it is done usually in the the field-free case. 

To finalize F-CT-MQC equations using the Floquet diabatic states, we give the explicit expression of the classical force of Eq.~(\ref{eqn: classical force 1}) for the trajectory $\nu$. In particular, we can compute the gradient of $\epsilon_{\mathrm{ext}}^\nu(t)$ from Eq.~(\ref{eqn: ext TDPES with states}), and the time derivative of the vector potential. Therefore, as in Eq.~(\ref{eqn: dot Csd 3}), we identify three contributions to the classical force,
\begin{align}\label{eqn: cl force}
\mathbf F_\nu(t)=\mathbf F_\nu^{\mathrm{EH}}(t)+\mathbf F_\nu^{\mathrm{QM}}(t)+\mathbf F_\nu^{\mathrm{EXT}}(t)
\end{align}
where `EH' and `QM' indicate the -- standard -- Ehrenfest-like~\cite{Agostini_CTC2019} and quantum-momentum terms, whereas `EXT' stands for the term depending on the external drive. The Ehrenfest-like contribution is 
\begin{align}
\mathbf F_\nu^{\mathrm{EH}}(t)&=\sum_{k,n}\left(-\boldsymbol{\nabla}E_{k}^\nu\right) \left|C_{k,n}^\nu(t)\right|^2+\sum_{k,n}\sum_{l,m}\bar C_{l,m}^\nu(t)C_{k,n}^\nu(t)\left(E_l^\nu-E_k^\nu\right)\mathbf d_{lk}^\nu\delta_{mn}
\end{align}
the quantum-momentum contribution, that couples the trajectories, is
\begin{align}
\mathbf F_\nu^{\mathrm{QM}}(t)&=\frac{2}{\hbar}\sum_{k,n}\left|C_{k,n}^\nu(t)\right|^2\left(\mathbf M^{-1}\boldsymbol{\mathcal P}^\nu(t)\cdot \mathbf f_{k,n}^\nu\right)\Big(\mathbf f_{k,n}^\nu-\mathbf A^\nu(t)\Big)
\end{align}
and the part that depends on the external field is
\begin{align}
\mathbf F_\nu^{\mathrm{EXT}}(t)&=\sum_{k,n}\sum_{l,m}\left(-\boldsymbol{\nabla}V_{mn,lk}^\nu\right)\bar C_{l,m}^\nu(t)C_{k,n}^\nu(t)+\frac{1}{\hbar}\sum_{k,n}\sum_{l,m}\mathrm{Im}\left[\bar C_{l,m}^\nu(t)C_{k,n}^\nu(t)\right]V_{mn,lk}^\nu\left(\mathbf f_{k,n}^\nu-\mathbf f_{l,m}^\nu\right)
\end{align}
Note that the last term in the definition of $\mathbf F_\nu^{\mathrm{EXT}}(t)$ is formally similar to the contribution due to spin-orbit coupling recently introduced in G-CT-MQC~\cite{Agostini_PRL2020, Agostini_JCTC2020_1}.

Before concluding this section, it is worth mentioning the partial normalization condition on the electronic wavefunction that is used to derive Eqs.~(\ref{eqn: EF n}) and~(\ref{eqn: EF el}). The partial normalization condition still holds, and using the expansion in harmonic-electronic eigenstates of the field-free Floquet Hamiltonian it reads 
\begin{subequations}
\begin{align}
1&=\frac{1}{T}\int_0^T ds\int d\mathbf r\sum_{k,n}\sum_{l,m} \bar C_{k,n}(\mathbf R,t)C_{l,m}(\mathbf R,t)e^{i(\omega_m-\omega_n)s} \bar{\psi}_k(\mathbf r;\mathbf R) \psi_l(\mathbf r;\mathbf R) \\
&= \sum_{k,n}\left|C_{k,n}(\mathbf R,t)\right|^2\quad \forall\,\, \mathbf R,t.
\end{align}
\end{subequations}
Equations~(\ref{eqn: R dot}), (\ref{eqn: dot Csd 3}) and~(\ref{eqn: cl force}) define the F-CT-MQC algorithm in the Floquet representation, that will be used in the following section to study the dynamics of a two-electronic-state one-dimensional model system subject to a cw laser of different intensities.

\section{Numerical studies}\label{sec: results}
The one-dimensional model employed for the numerical studies using F-CT-MQC is defined by the field-free electronic Hamiltonian 
\begin{align}\label{eqn: model Hel}
\hat H_{el}= \left(
\begin{array}{cc}
H_{11}(R) & H_{12}(R) \\
H_{12}(R) & H_{22}(R)
\end{array}
\right) = 
\left(
\begin{array}{cc}
\frac{1}{2}K(R-R_1)^2 & \gamma e^{-\alpha(R-R_3)^2} \\
\gamma e^{-\alpha(R-R_3)^2} & \frac{1}{2}K(R-R_2)^2+\Delta
\end{array}
\right)
\end{align}
given in the diabatic basis and depending on one nuclear coordinate $R$. The parameters are chosen as: $K=0.02$~au, $\Delta=0.01$~au, $\gamma=0.01$~au, $\alpha=3.0$~au, $R_1=6.0$~au, $R_2=2.0$~au, $R_3=3.875$~au. The diabatic PESs are parabolas, one with minimum in $R_1$ and the other with minimum in $R_2<R_1$; the parabolas are, thus, displaced in $R$, and in energy as well by the amount $\Delta$; the diabatic coupling has a Gaussian shape, with maximum in $R_3$, and strength regulated by the parameter $\gamma$. The energy difference between the minimum of the $H_{22}(R)$ potential, at $R=R_2$~au, where the initial wavepacket will be prepared for the dynamics described in the following, and $H_{11}(R_2)$ is $\Delta E = 0.15$~au.

The interaction with the laser field is described in the dipole approximation as $\hat V(t) =-\hat\mu E(t) = -\hat\mu E_0\cos(\Omega t)$, with $\hat\mu$ the dipole moment operator. Neglecting the nuclear contribution to the total dipole moment, and setting to zero the diagonal elements of $\hat\mu$ in the diabatic basis, the off-diagonal elements of the system-laser interaction Hamiltonian are $-\mu(R)E_0\cos(\Omega t)=-\beta RE_0\cos(\Omega t)$.
The transition dipole moment $\mu(R)=\beta R$ is chosen to be a linear function of $R$, and $\beta=0.05$~au. The frequency of the driving field is set as $\Omega=0.05$~au, and two cases will be studied to test F-CT-MQC, a weak field case with $E_0=0.25$~au 
and a strong field case with $E_0=0.5$~au.

Quantum dynamics is initiated in the ground vibrational state of the left diabatic potential well (corresponding to $S_0$), centered at $R=2$~au with average zero initial momentum. Nuclear mass has been set to the value $M=20000$~au. Reference results are obtained by the integrating the full tdSE in the diabatic basis by using the split-operator technique~\cite{spo} with a time step of $dt=0.1$~au. Since the Hamiltonian explicitly depends on time, the stability of the numerical integration has been confirmed by monitoring the norm of the wavefunction. The diabatic-to-adiabatic change of basis has been performed to compute the population of the adiabatic states $S_0$ and $S_1$, for comparison with F-CT-MQC calculations. While quantum dynamics is performed to determine the full wavefunction as function of time, its exact-factorization form can be determined as well, in order to compute the TDPES (that will be shown below). To this end, a choice of gauge has to be made. In the quantum dynamics case, the gauge can be set by imposing that the time-dependent vector potential is zero.

Trajectory-based F-CT-MQC calculations are performed in the Floquet diabatic basis. $N_{tr}=100$ trajectories are evolved according to Eqs.~(\ref{eqn: R dot}) and~(\ref{eqn: cl force}) with the velocity-Verlet algorithm starting with Wigner-sampled initial conditions (the Wigner distribution is determined as the Wigner transform of the initial nuclear probability density of quantum calculations). The Fourier series used to represent the Floquet eigenmodes is truncated to include $2 N_{max}+1$ components. In the calculations presented below, convergence is tested by increasing the size of the basis with $N_{max}=1,2,3,4,5$ (calculations with $N_{max}=6,7$ have been performed as well but they are not shown because convergence is reached with $N_{max}=5$). The electronic initial condition is chosen by setting equal to one the harmonic-electronic coefficient corresponding to $k=S_0$ and $n=0$ for all trajectories. We recall that the index $k$ indicates the electronic state, in this case $k$ can be $S_0$ or $S_1$, and $n$ indicates the harmonic, in this case $n\in[-N_{max},N_{max}]$. With this initial condition, the electronic equation~(\ref{eqn: dot Csd 3}) is integrated with the forth-order Runge-Kutta algorithm. The time step for the nuclear and the electronic integration is $dt=0.1$~au.

F-CT-MQC calculations provide direct access to the populations of the Floquet diabatic states along a trajectory $\nu$. In order to compare trajectory-based results to the reference results, the populations $\rho_k(t)$ of the adiabatic states $k=S_0,S_1$ as functions of time are determined as
\begin{align}\label{eqn: BO pop}
\rho_k(t) = \frac{1}{N_{tr}} \sum_{\nu=1}^{N_{tr}} \left|\sum_{n=-N_{max}}^{n=N_{max}}C_{k,n}^\nu(t)e^{i\omega_nt}\right|^2
\end{align}
Similarly, it is possible to interpret the dynamics in terms of the number of photons exchanged between the system and the external field. To this end, the photon population can be estimated as
\begin{align}\label{eqn: photon pop}
\rho_n(t) = \frac{1}{N_{tr}} \sum_{\nu=1}^{N_{tr}}\sum_{k=S_0,S_1} \left|C_{k,n}^\nu(t)\right|^2
\end{align}
where $n\in[-N_{max},N_{max}]$.

\subsection{Case of a weak cw laser}
\begin{figure}[h!]
\centering
\includegraphics[width=.8\textwidth]{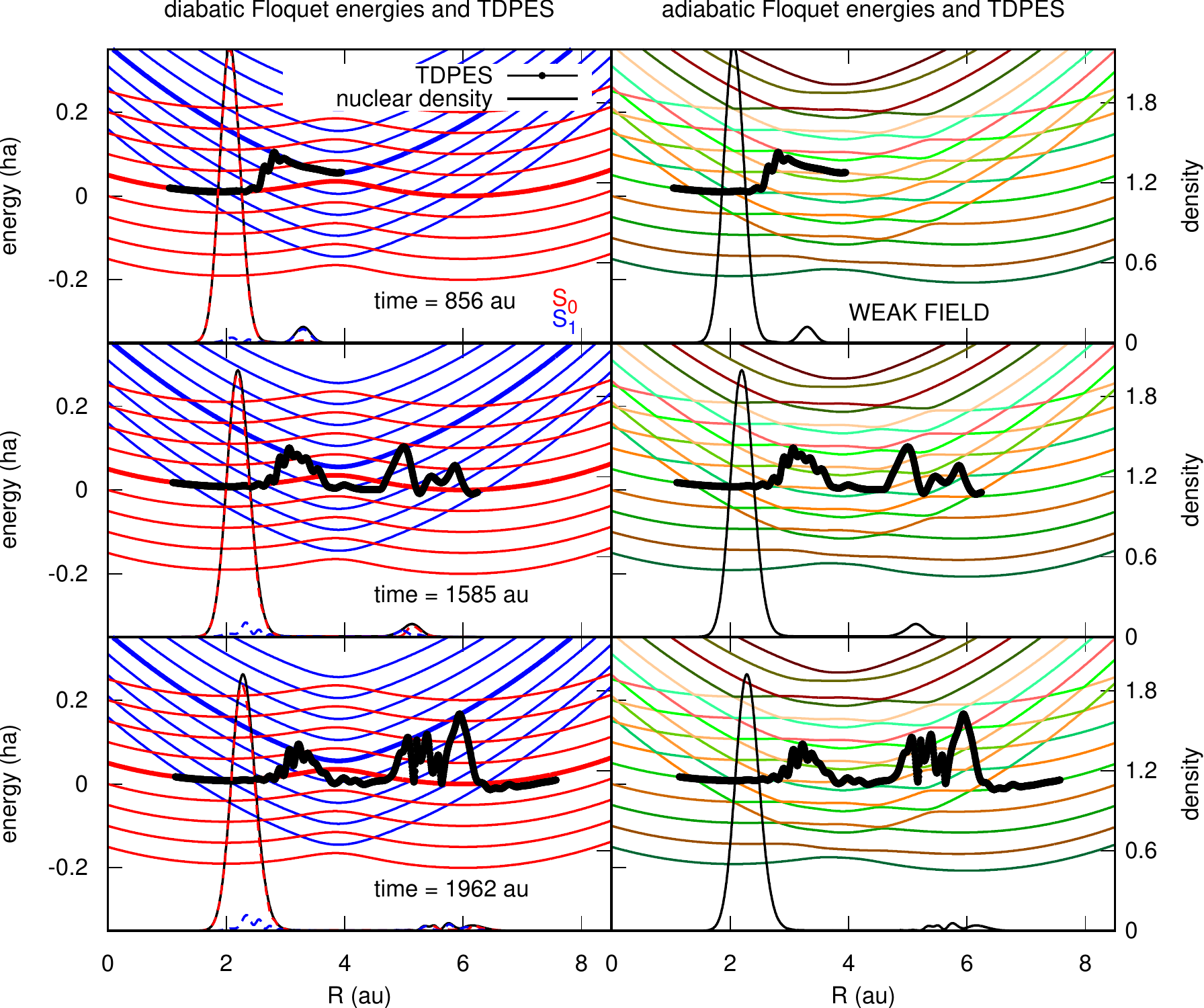}
\caption{Weak-field case. Left panels: Floquet diabatic PESs ($S_0$ red lines and $S_1$ blue lines) with $N_{max}=4$. The thick red and blue lines identify the $S_0$ and $S_1$ PESs for the harmonic $n=0$. Nuclear density (black thin lines), along with its decomposition in $S_0$ (dashed red lines) and $S_1$ (dashed blue lines) contributions, and the gauge-invariant part of the TDPES (black line-dots) are shown at times $t=856,1585,1962$~au. Right panels: Floquet adiabatic PESs computed with $N_{max}=4$. Nuclear density (black thin lines) and the TDPES (black line-dots) are shown at the same times as in the left panels.}
\label{fig: TDPES weak laser}
\end{figure}

In the weak-field case, some of the density that is initially prepared in the vibrational ground state of the left well, in $S_0$, is transferred to the electronic excited state $S_1$. While the field keeps driving population back and forth between the two states, with the main portion of the nuclear density still centered around $R=2$~au in $S_0$, the excited wavepacket evolves towards the avoided crossing located at $R\sim 4$~au, and transfers back population to the ground state. The dynamics just described is detailed in Fig.~\ref{fig: TDPES weak laser} where the nuclear density at three different times is shown, along with its decomposition in terms of Born-Oppenheimer contributions and the gauge-invariant part of the TDPES given as the sum of Eqs.~(\ref{eqn: TDPES}) and~(\ref{eqn: TDPES ext}) (left panels).

The gauge-invariant part of the TDPES does not contain the term with the partial time derivative of Eq.~(\ref{eqn: TDPES}). It has been shown in previous work~\cite{Gross_MP2013, Gross_JCP2015, Curchod_JCP2016}, and it is the case here as well, that this gauge-dependent contribution to the TDPES nearly mirrors the shape of the gauge-invariant part, and it is a piecewise constant function of $R$; its main effect is to reduce the height of the steps, as the step observed at $t=856$~au in Fig.~\ref{fig: TDPES weak laser} at about $R=3$~au, but leaving unchanged the slope of the gauge-invariant TDPES outside the steps. It should be noted that, in the present case, where an external time-dependent field is applied, the features of the gauge-invariant and gauge-dependent parts of the TDPES are not as neat as in the field-free case. 

The TDPES provides information about the energy scale involved in the studied process, and, consequently, its comparison with the Floquet (a)diabatic PESs allows us to estimate the number of harmonics required to achieve convergence of the Floquet-based calculations. With this comparison in mind, we show in Fig.~\ref{fig: TDPES weak laser} the Floquet diabatic PESs (left panel) and the Floquet adiabatic PESs (right panels). Floquet diabatic PESs are simply the Born-Oppenheimer PESs, obtained by  diagonalization of the electronic Hamiltonian~(\ref{eqn: model Hel}) at each $R$, shifted by the constant energy of the harmonics $n\hbar\Omega$. In Fig.~\ref{fig: TDPES weak laser}, the case $N_{max}=4$ is shown, and the thick red and blue lines refer to the harmonic $n=0$. In this case, all along the dynamics the TDPES oscillates within energy values comprised between $E_{S_0}(R)-4\hbar\Omega$ and $E_{S_1}(R)+4\hbar\Omega$, suggesting that F-CT-MQC results should converge with $N_{max}=4$.

On the right panels of Fig.~\ref{fig: TDPES weak laser}, nuclear dynamics and the TDPES are superimposed to the Floquet adiabatic PESs, which are the eigenvalues of the Floquet Hamiltonian explicitly including the external drive. Also in this case, the Floquet basis is truncated to $N_{max}=4$. The main observation arising from the comparison of the left panels and right panels of Fig.~\ref{fig: TDPES weak laser} is that the Floquet adiabatic PESs can be very different from the Floquet diabatic PESs, as they represent the energy of hybrid electronic and field states that are strongly coupled to each other. Additional observations on the relation between Floquet adiabatic and Floquet diabatic PESs is reported in Section~\ref{sec: observations}.

\begin{figure}
\centering
\includegraphics[width=.6\textwidth]{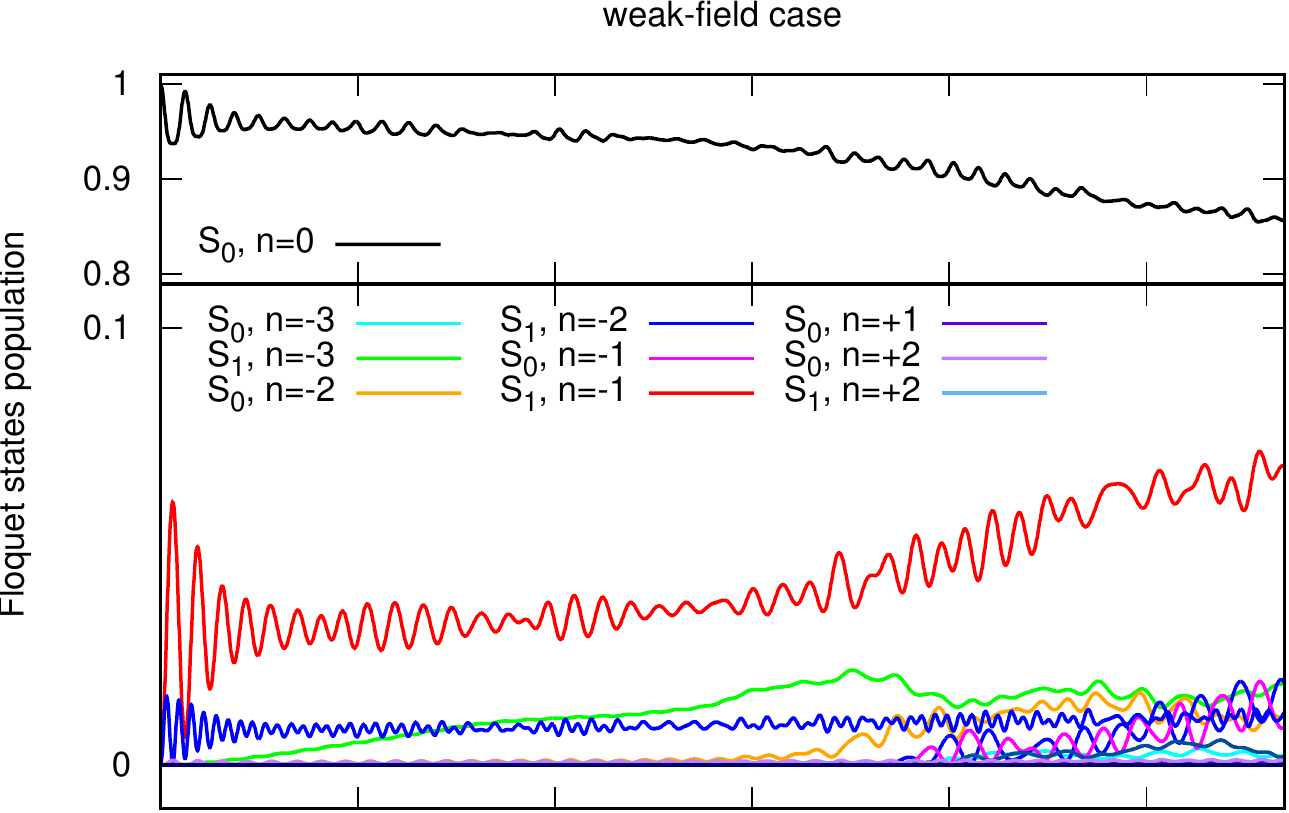}
\caption{Weak-field case. Populations of the Floquet diabatic states as functions of time. Only the non-zero populations are shown.}
\label{fig: Floquet pop weak laser}
\end{figure}

Figure~\ref{fig: Floquet pop weak laser} shows the populations of the Floquet diabatic states for the simulation where $N_{max}=5$. These quantities are estimated as the average over the trajectories of the squared moduli of the coefficients $C_{k,n}^\nu(t)$. The analysis of the populations reported in Fig.~\ref{fig: Floquet pop weak laser} gives information about the number of photons exchanged between the system and the external field during the dynamical process.
Figure~\ref{fig: Floquet pop weak laser} shows that only few Floquet diabatic states are substantially populated during the dynamics. The state labelled $k=S_0,n=0$ starts with full occupation, but rapidly transfers some population to $k=S_1,n=-1$, suggesting that a one-photon process is taking place to excite a portion of the ground-state wavepacket. Since the nonadiabatic couplings mediate population transfer between different electronic states within the same harmonic, and since in the region where the nuclear density is prepared these couplings are mostly zero, population transfer at the initial times is only driven by the external field. A small amount of population is also transferred to $k=S_1,n=+1$ at the initial time, once again suggesting a one-photon process, but the population remains close to zero ($<0.02 \,\forall\,t$), as well as all other populations but $k=S_0,n=0$ and $k=S_1,n=-1$. 

\begin{figure}
\centering
\includegraphics[width=.6\textwidth]{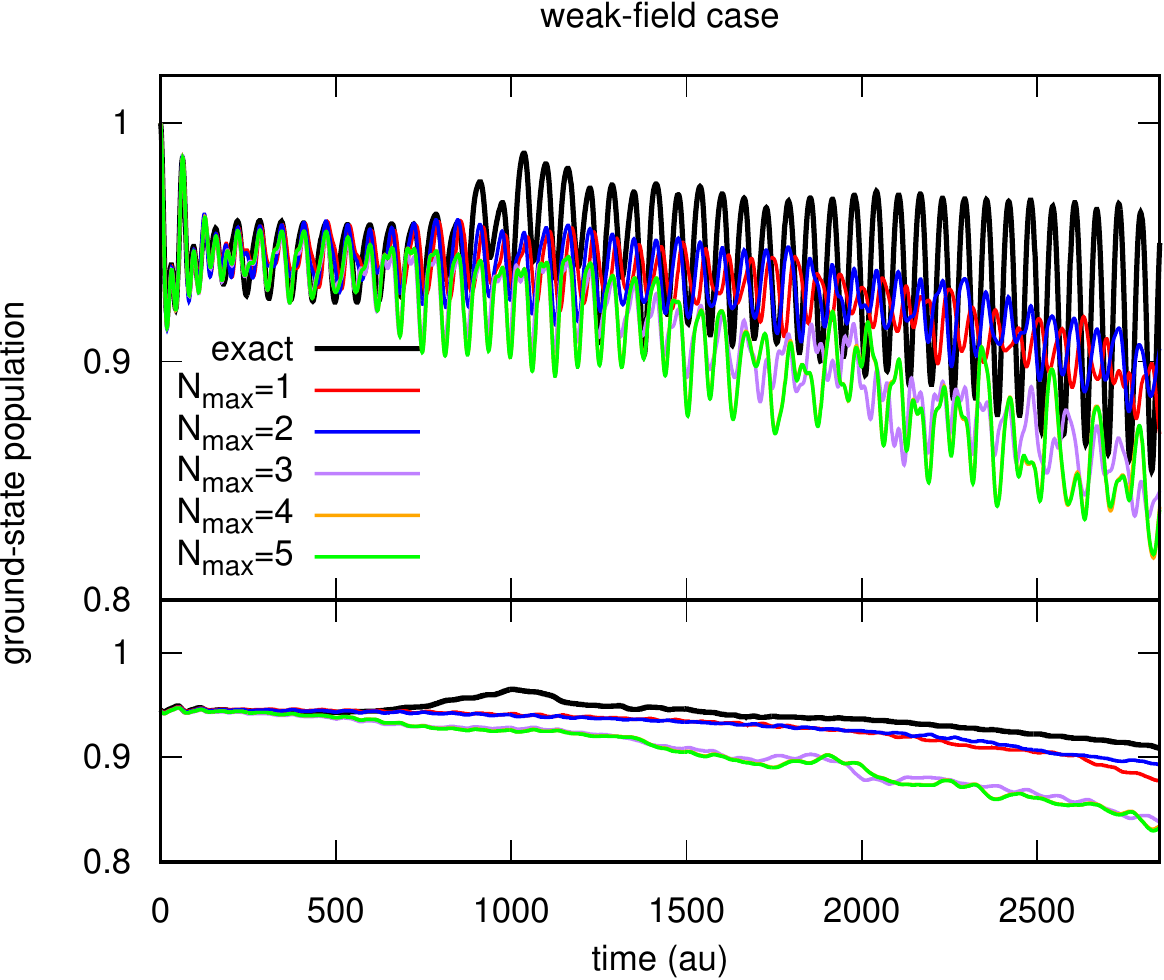}
\caption{Weak-field case. Upper panel: population of the $S_0$ state as function of time, from exact quantum-dynamics calculations (black line), and from F-CT-MQC calculations with $N_{max}=1$ (red line), $N_{max}=2$ (blue line), $N_{max}=3$ (purple line), $N_{max}=4$ (orange line), $N_{max}=5$ (green line). Lower panel: population of the $S_0$ state as function of time, averaged over a period of the driving field with Eq.~(\ref{eqn: average BO pop}), from exact quantum-dynamics calculations, and from F-CT-MQC calculations with $N_{max}=1,2,3,4,5$. The colore code is the same as in the upper panel.}
\label{fig: pop weak laser}
\end{figure}
Using Eq.~(\ref{eqn: BO pop}), we can estimate the populations of the physical electronic states as functions of time from the Floquet-based calculations. The population of the ground state $S_0$ is shown in Fig.~\ref{fig: pop weak laser}. In the upper panel, the estimate from Eq.~(\ref{eqn: BO pop}) with $k=S_0$ is used, with different values of $N_{max}$ for F-CT-MQC calculations, to test convergence with the number of harmonics by comparing with exact results. In the lower panel, the oscillations in the populations are averaged out by computing a moving average over a period, namely 
\begin{align}\label{eqn: average BO pop}
\langle\rho_k(t)\rangle_T = \frac{1}{T}\int_t^{t+T} d\tau\frac{1}{N_{tr}} \sum_{\nu=1}^{N_{tr}} \left|\sum_{n=-N_{max}}^{n=N_{max}}C_{k,n}^\nu(\tau)e^{i\omega_n\tau}\right|^2
\end{align}
and similarly for quantum calculations, which allows us to focus on the overall behavior of the population as function of time. As predicted from the analysis of Fig.~\ref{fig: TDPES weak laser} of the TDPES, $N_{max}=4$ is sufficient to converge F-CT-MQC, even though not to the exact value of the $S_0$ population. However, for the simulated dynamics, the relative error on the population remains within 10\% of the reference (estimated from the lower panel of Fig.~\ref{fig: pop weak laser}). Finally, note that, in the upper panel of Fig.~\ref{fig: pop weak laser}, the amplitude of the oscillations of F-CT-MQC results is smaller if compared to exact results, probably as effect of the average operation over the trajectories.

\begin{figure}
\centering
\includegraphics[width=.6\textwidth]{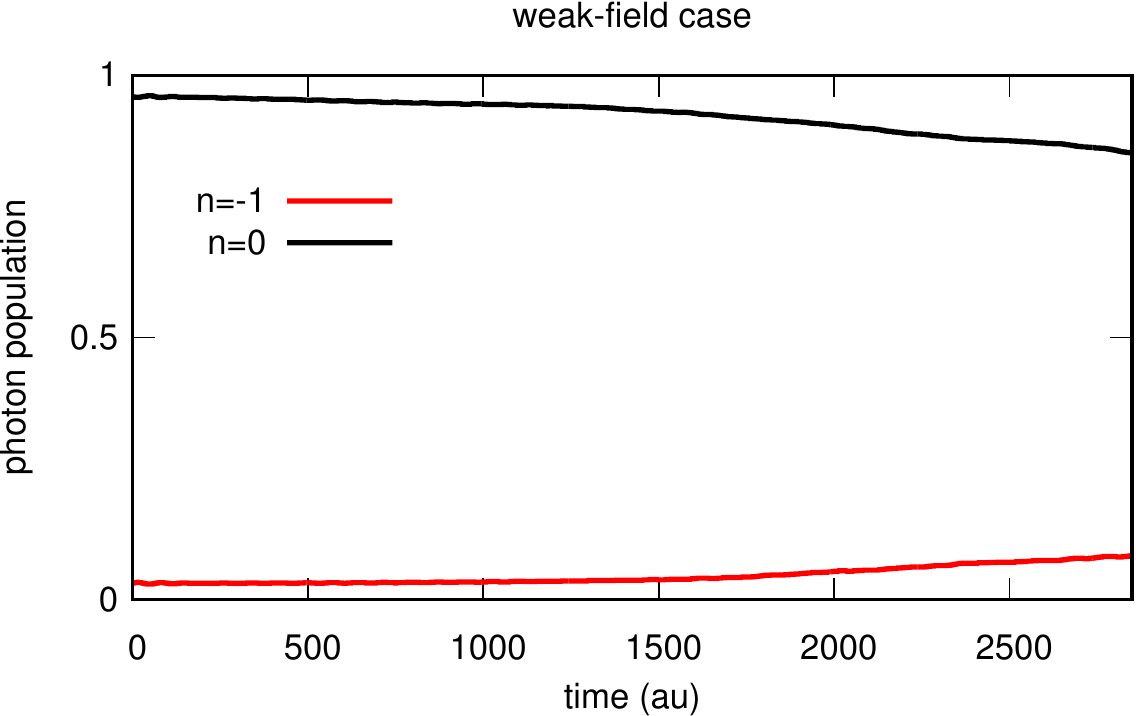}
\caption{Weak-field case. Photon population as function of time, estimated from Eq.~(\ref{eqn: photon pop}) and averaged over a period of the driving field with an expression equivalent to Eq.~(\ref{eqn: average BO pop}). Only the populations with values $> 0.05$ $\forall\, t$ are shown.}
\label{fig: Nph weak laser}
\end{figure}

Using Eq.~(\ref{eqn: photon pop}), we can estimate the photon population to interpret the dynamics in terms of single- or multi-photon absorption and emission processes. In order to average out the fast oscillations in photon populations that are similar to the oscillations observed in the upper panel of Fig.~\ref{fig: pop weak laser}, we show in Fig.~\ref{fig: Nph weak laser} $\langle\rho_n(t)\rangle_T$, which is a quantity analogous to the one derived in Eq.~(\ref{eqn: average BO pop}), but applied to the photon population.
Figure~\ref{fig: Nph weak laser} shows that the dynamics is dominated by zero-photon processes, with a small contribution from one-photon absorption (negative values of $n$) mainly appearing after 2000~au.

\begin{figure}
\centering
\includegraphics[width=.8\textwidth]{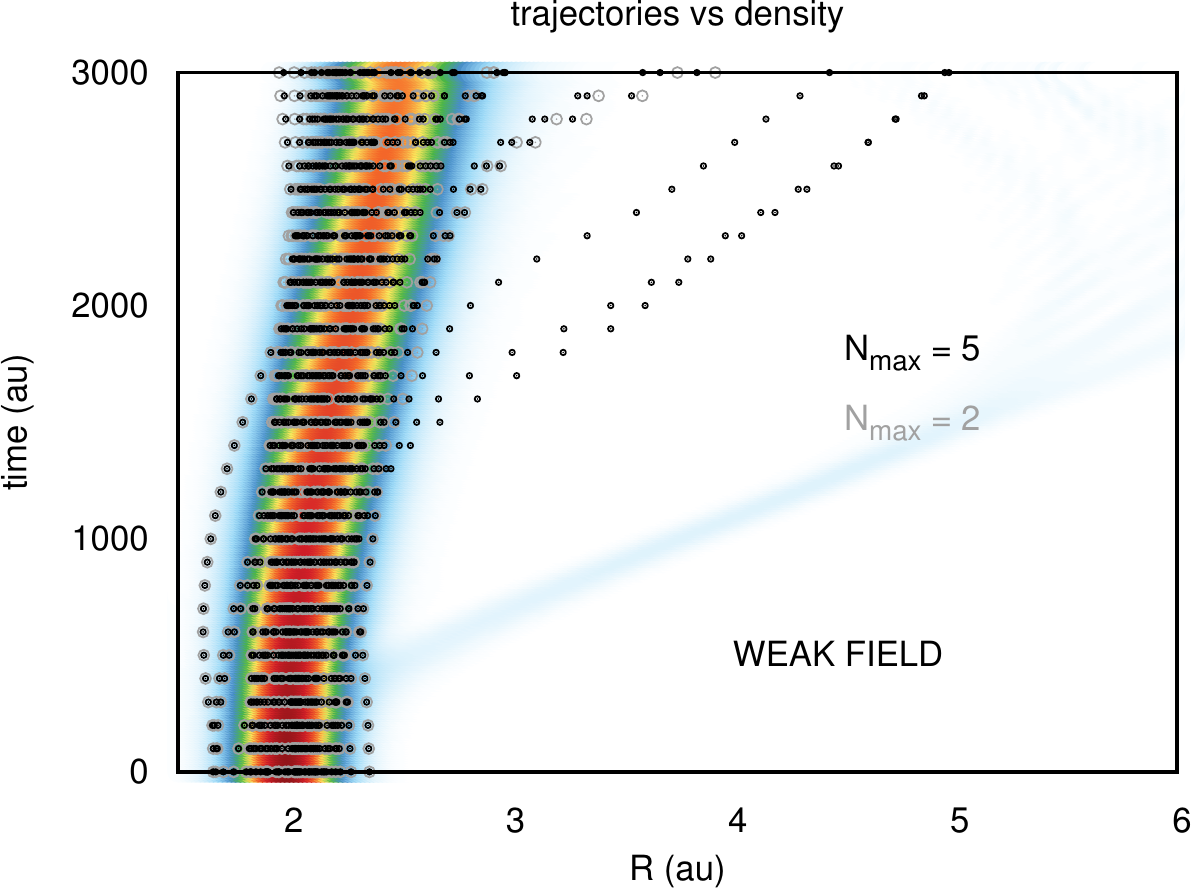}
\caption{Weak-field case. Colored areas: nuclear density as function of $R$ and $t$. Circle: distributions of classical trajectories for $N_{max}=2$ (gray circle) and $N_{max}=5$ (black circles) plotted every 100~au.}
\label{fig: density weak laser}
\end{figure}

In Fig.~\ref{fig: density weak laser}, we compare the nuclear density as function of time with the distribution of F-CT-MQC trajectories for $N_{max}=2$ (gray circles) and $N_{max}=5$ (black circles). As discussed above, the nuclear density mainly remains localized around $R=2$~au at all times, even though already at about 500~au, a small portion of the density -- evolving on the excited state -- moves towards large values of $R$. With $N_{max}=2$, classical trajectories follow closely the main portion of the nuclear density, but the diverging branch is missed. With $N_{max}=5$, instead, some trajectories diverge from the main bundle localized around $R=2$~au, even though the time scale is not as in the reference results, and the trajectories do not go as far as the quantum wavepacket within the simulated dynamics. 

\subsection{Case of a strong cw laser}
\begin{figure}
\centering
\includegraphics[width=.8\textwidth]{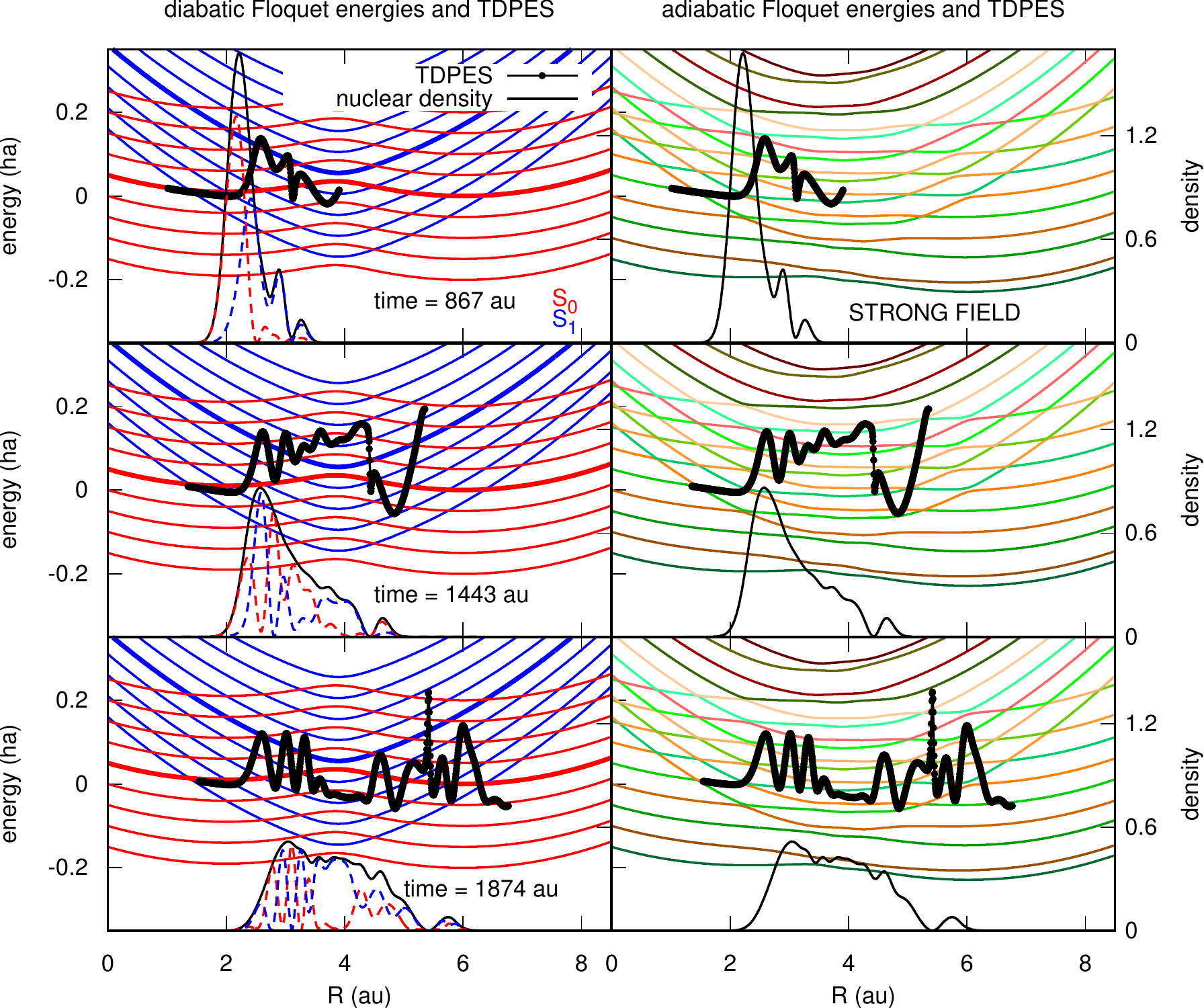}
\caption{Strong-field case. The panels are similar to those in Fig.~\ref{fig: TDPES weak laser} and the same color-code is used. The nuclear density and the TDPES are shown at times $t=867,1443,1874$~au.}
\label{fig: TDPES strong laser}
\end{figure}

Even though the amplitude of the external field is only doubled in going from the weak-field case to the strong-field case, the dynamics in this second example is very different with respect to the previous one, which can be confirmed by comparing nuclear dynamics from Fig.~\ref{fig: TDPES weak laser} and from Fig.~\ref{fig: TDPES strong laser}. However, similar general conclusions on the performance of F-CT-MQC can be drawn.

Let us first analyze the dynamics shown in Fig.~\ref{fig: TDPES strong laser}, based on the comparison between the TDPES and the Floquet diabatic PESs (left panels) or the Floquet adiabatic PESs (right panels). The TDPES is almost at all times comprised within the Floquet diabatic PESs shown in the figure, for $N_{max}=4$.
In this case, the external field excites the nuclear wavepacket from $S_0$ to $S_1$. While population transfer between the two states is continuously driven, the excited portion of the wavepacket moves towards the avoided crossing and transfer population back to the ground state. At the same time, the ground-state wavepacket is driven towards the avoided crossing itself, where it encounters the excited-state wavepacket, producing complex interference patterns at long times. Clearly, this dynamics is very difficult to capture based on classical trajectories without accounting for interference effects.

\begin{figure}
\centering
\includegraphics[width=.8\textwidth]{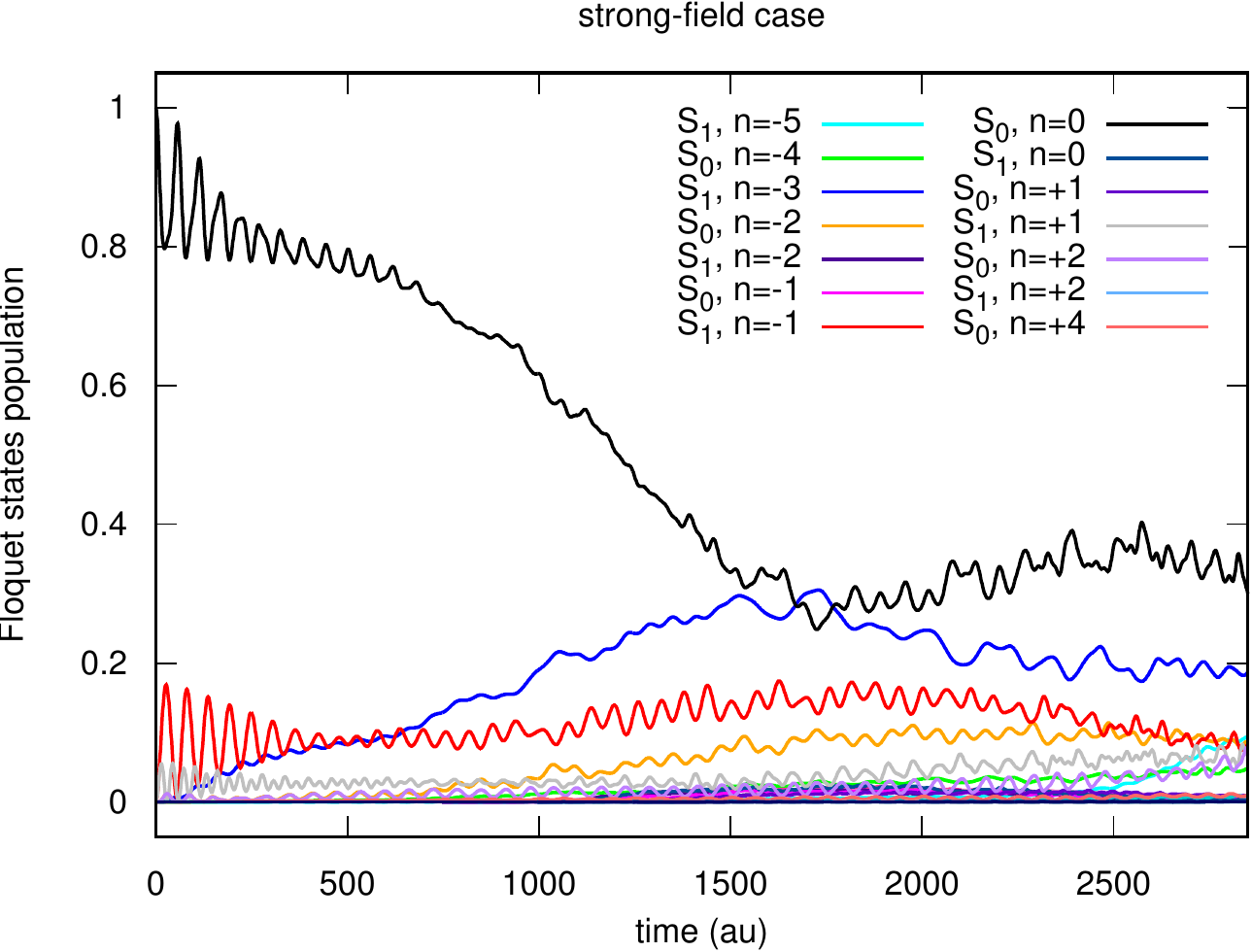}
\caption{Strong-field case. Populations of the Floquet diabatic states as functions of time. Only the non-zero populations are shown.}
\label{fig: Floquet pop strong laser}
\end{figure}
Analysis of the Floquet diabatic populations shows that multi-photon processes take place during the dynamics. As in the previous case, the state labelled $k=S_0,n=0$ is fully populated initially, but it rapidly transfers population mainly to $k=S_1,n=-3$, to $k=S_1,n=-1$ and to $k=S_1,n=+1$ . At later times, after 1000~au, other states become populated, even if their population remains below 0.1. 

\begin{figure}
\centering
\includegraphics[width=.6\textwidth]{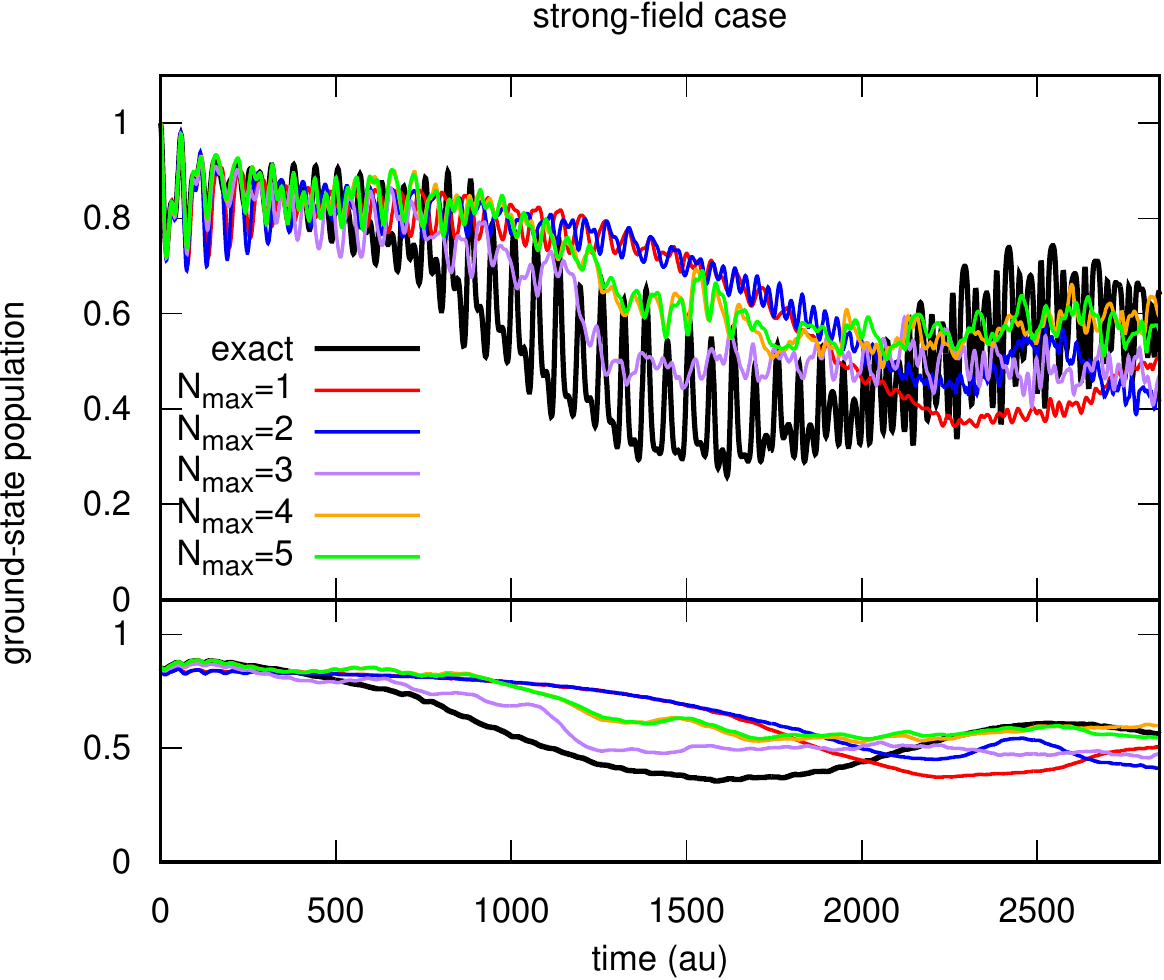}
\caption{Strong-field case. The panels are similar to those in Fig.~\ref{fig: pop weak laser} and the same color-code is used.}
\label{fig: pop strong laser}
\end{figure}

Applying Eq.~(\ref{eqn: BO pop}), the populations of $S_0$ and $S_1$ states are determined from the populations of the Floquet diabatic states, and the $S_0$ population is shown in Fig.~\ref{fig: pop strong laser}. As mentioned above, convergence in achieved for $N_{max}=4$, but the results slightly differ from the reference calculations. In particular, the low frequency oscillations of the $S_0$ population are not quite captured, as it is evident in the plot of the populations averaged over a period of the drive (lower panel). Indeed, inclusion of more harmonics, for instance from $N_{max}=2$ to $N_{max}=4$, allows to better reproduce the decay between 500~au and 1500~au, even though quantitative agreement is missing all along the simulated dynamics.
Also in this case, and probably stronger than in the weak-field case, the amplitude of the high-frequency oscillations of the population (upper panel) is suppressed in F-CT-MQC calculations, once again as effect of the average over the trajectories.

\begin{figure}
\centering
\includegraphics[width=.6\textwidth]{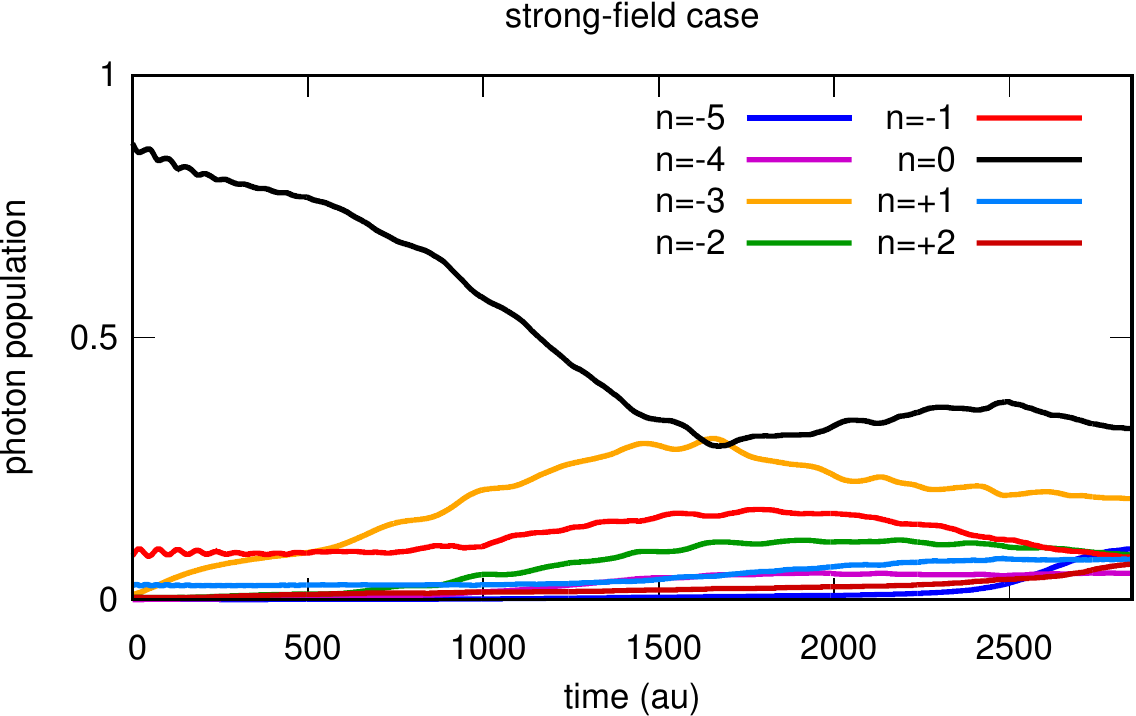}
\caption{Strong-field case. Photon population as function of time, estimated from Eq.~(\ref{eqn: photon pop}) and averaged over a period of the driving field with an expression equivalent to Eq.~(\ref{eqn: average BO pop}). Only the populations with values $> 0.05$ $\forall\, t$ are shown.}
\label{fig: Nph strong laser}
\end{figure}

Similarly to the weak-field case, the populations of the Floquet states allows for the calculation of photon populations. Figure~\ref{fig: Nph strong laser} shows that at short times the dynamics is dominated by a zero-photon process, since the population $\rho_{n=0}(t)$ is much larger that other contributions for $t<1000$~au. By contrast with the weak-field case, however, a single-photon absorption process has non-zero contribution since the beginning of the simulated dynamics. After 1000~au, however, multi-photon absorption processes are observed, as well as single- and two-photon emissions ($\rho_{n=+1}(t)$ and $\rho_{n=+2}(t)$ are small but non-zero after 2000~au).

\begin{figure}
\centering
\includegraphics[width=.8\textwidth]{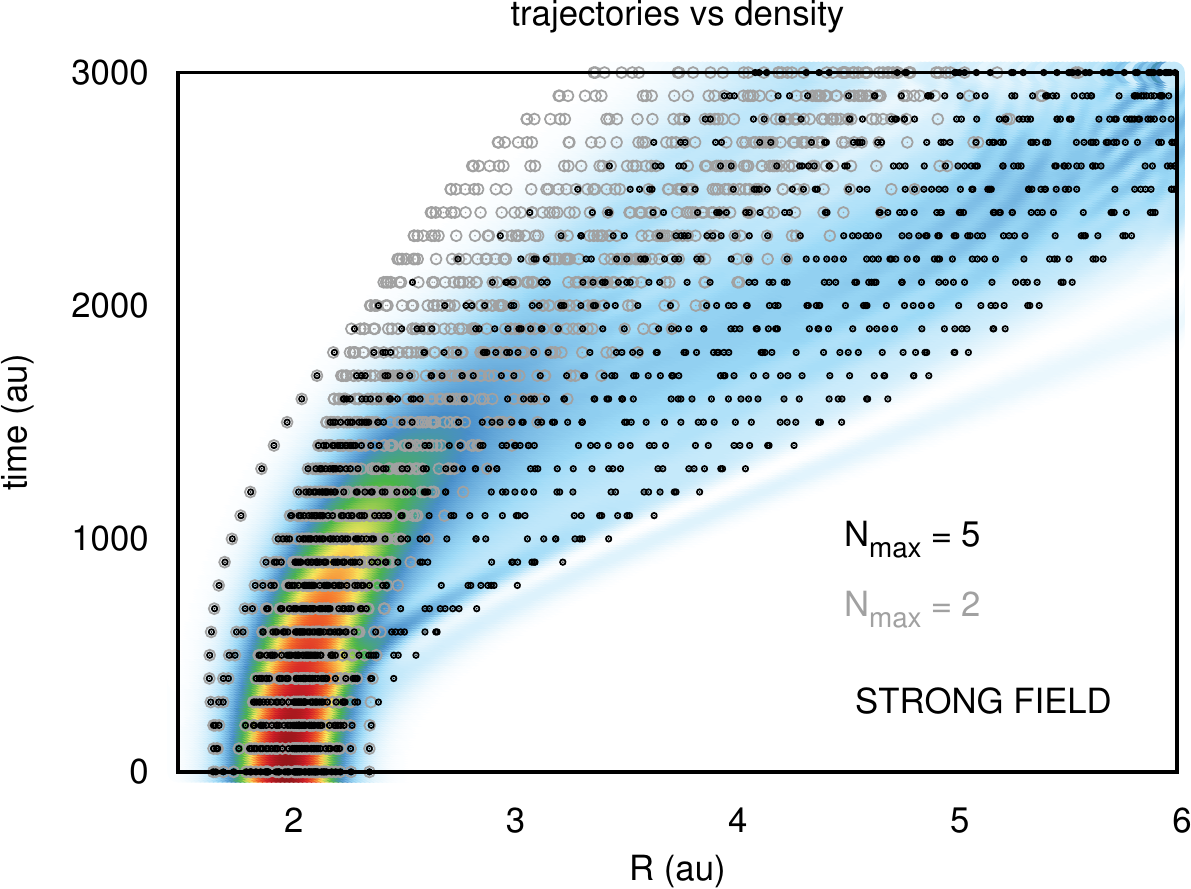}
\caption{Strong-field case. Similar to the results reported in Fig.~\ref{fig: density weak laser} with the same color-code.}
\label{fig: density strong laser}
\end{figure}
Finally, we compare in Fig.~\ref{fig: density strong laser} the nuclear probability density with the distribution of F-CT-MQC trajectories for $N_{max}=2$ (gray circles) and $N_{max}=5$ (black circles). In general, we observe that the density spreads over time and moves towards large values of $R$. For $N_{max}=2$ the trajectories only follow one portion of the nuclear density, whereas for $N_{max}=5$ the whole nuclear configuration space spanned by the quantum wavepacket is sampled as well by the trajectories.

In this case, and in contrast with the weak-field case, the maximum number of harmonics included in F-CT-MQC calculations provides satisfactory results for nuclear dynamics, while it misses some features in the electronic populations. Conversely, in the weak-field case, nuclear dynamics was not correctly captured by the trajectories, but a better agreement was achieved for the electronic populations. 

\subsection{General discussion on F-CT-MQC}\label{sec: observations}
The exact factorization, together with its trajectory-based treatment, can be formulated employing the Floquet formalism in different, equivalent, ways. In Section~\ref{sec: theory}, we have already discussed two possibilities, which we referred to as Floquet adiabatic and Floquet diabatic picture. Nonetheless, the explicit derivation of F-CT-MQC equations and the numerical results have been presented only in the Floquet diabatic representation. The choice has been made based on practical reasons: (i) the preparation of the initial state for the dynamics, and (ii) the calculation of energy gradients and derivative couplings. The initial state of a simulation is usually chosen to be one of the ``physical''  states of the system, e.g., the ground state, which is easily identified in the Floquet diabatic picture by selecting the harmonic $n=0$. However, in the Floquet adiabatic picture, this identification is not straightforward. In addition, quantum-chemistry codes usually provide gradients of the Born-Oppenheimer PESs, which are identical to the gradients of Floquet diabatic PESs, but not of Floquet adiabatic PESs. This is clearly a limitation for the possibility of exploiting a trajectory-based algorithm that employs the Floquet adiabatic representation in combination with quantum chemistry for molecular calculations. We should also mention that some Floquet adiabatic PESs present trivial crossings, where the PESs are degenerate and, thus, the derivative couplings are singular. Trivial crossings need special treatment in combination with trajectory-based simulations, and clearly depend on the particular form of the external field. All these issues are easily circumvented in the Floquet diabatic picture. However, an intriguing question remains, as to whether electronic-structure properties in the Floquet adiabatic representation are easily accessible based on standard quantum-chemistry theory (and codes).

It is important to mention that, F-CT-MQC is readily suitable for on-the-fly calculations in combination with ab initio PESs and derivative couplings. The only additional elements used in Section~\ref{sec: results} in comparison to field-free simulations are the electronic transition dipole moment of the molecule and its spatial derivatives, which have to be computed at each nuclear configuration visited by the trajectories, and are both accessible based on quantum-chemistry calculations~\cite{Martinez_PCCP2017}.

In the literature~\cite{Subotnik_JCP2020, Subotnik_JCTC2020, Schmidt_PRA2016, Gonzalez_JPCA2012}, the Floquet diabatic picture has been preferred over the Floquet adiabatic one, especially in combination with the trajectory surface hopping algorithm. In fact, as shown in Figs.~\ref{fig: TDPES weak laser} and~\ref{fig: TDPES strong laser},  the Floquet adiabatic PESs can present many avoided crossings, as well as trivial crossings. These features potentially pose challenges for the fewest switches surface hopping algorithm, because the trajectories hop often from one state to another; in addition, a large number of trajectories is required for convergence in order to sample satisfactorily the ``hopping space''. Perhaps, when working with dense manifolds of coupled Floquet adiabatic states, an Ehrenfest-like approach would be preferable. This suggests that F-CT-MQC in the Floquet adiabatic representation might perform well, similarly to the case of Floquet diabatic representation, due to its Ehrenfest-like form -- provided that the issues (i) and (ii) above are efficiently circumvented.

An alternative strategy for deriving F-CT-MQC can be envisaged: rather than invoking the Fourier representation of Floquet eigenmodes, and introducing the harmonics to account for the periodicity of the external drive in the electronic basis, one could directly work with Floquet eigenmodes, as, for instance, those determined by solving the eigenvalue problem~(\ref{eqn: Fl adiabatic eqn}). This avenue has not been explored in this work, but it would be interesting to investigate how the two formulations compare in the trajectory-based scheme. Clearly, if the exact equations were solved, the two formulations would yield the same result, however, this is not always true when approximations are invoked.

An additional approach to include the effect of the external drive has been tested, but the results have not been reported in the present work. Such an approach simply includes the external drive in CT-MQC equations, where the used electronic representation is the ``standard'' Born-Oppenheimer representation. CT-MQC yields, in this case, directly the populations of the physical electronic states, and interpretation in terms of photon absorption or emission processes is not possible. For the model studied here, trajectory-based results are very close to exact and Floquet-based results in both weak- and strong-field cases. Therefore, even though such strategy to include the external time-dependent field in CT-MQC has not been yet systematically investigated, it is a promising route for future theoretical and numerical developments.


\section{Conclusions}\label{sec: conclusions}
Trajectory-based excited-state simulations of systems subject to an external drive are nowadays very challenging. Standard trajectory-based schemes might yield different results depending on the used electronic basis, or electronic representation, since the approximations they are based on are derived in a particular electronic basis. The choice of electronic representation becomes, thus, of paramount importance. In order to easily generalize standard schemes, one might be tempted to use the so-called quasi-static representation, where the electronic states are defined at each time of the propagation as the eigenstates of the electronic problem for a given nuclear configuration (as in standard field-free situations) including the effect of the external time-dependent field. However, unless that external field varies slowly in time, the electronic quasi-static states are not \textsl{descriptive} of the state of the system, and numerical simulations yield unphysical results~\footnote{Note that, this strategy has been tested for the cases studied in the present paper. While the high-frequency oscillations of the adiabatic populations driven by the external field were captured in the quasi-static representation, no population transfer between the $S_0$ and $S_1$ states was observed for both field strengths studied in this work.}. The Floquet formalism is, instead, adequate to treat periodically driven systems because the periodic time dependence is somehow treated in a static way -- in Fourier space -- by extending the electronic space to the space of harmonics. The drawbacks are, clearly, the fact the electronic Hamiltonian becomes unbounded, and that convergence studies on the number of harmonics to be included have to be carried out (and depending on the intensity of the drive, a large number of harmonics need to be considered).

As it has been shown in previous work~\cite{Maitra_PRL2019, Tokatly_EPJB2018, Maitra_EPJB2018, Schmidt_PRA2017, Maitra_PCCP2017, Maitra_PRL2015, Gross_PRL2010, Suzuki_PCCP2015, Suzuki_PRA2014}, the exact-factorization formalism naturally lends itself to the treatment of dynamics in the presence of an external time-dependent field. In addition, the Floquet formalism can be used together with the exact factorization to interpret and to justify common approaches based on trajectories~\cite{Schmidt_PRA2017}. In the present work, we took a step forward, and we employed the trajectory-based solution of the exact-factorization equations, i.e., the CT-MQC scheme, together with the Floquet formalism, to propose a new algorithm designed to treat periodically driven electron-nuclear systems in the presence of nonadiabatic effects. The CT-MQC algorithm has the advantage of being easily adapted to treat various physical situations, like standard nonadiabatic effects~\cite{Gross_PRL2015, Gross_JCTC2016, Gross_JPCL2017}, spin-orbit coupling~\cite{Agostini_PRL2020, Agostini_JCTC2020_1} (G-CT-MQC), or external time-dependent fields (F-CT-MQC). The fundamental equations do not have to be completely modified or adapted to include different effects, since such additional effects can be easily included starting from the -- quantum-mechanical -- formulation of the time-dependent Schr\"odinger equation in its exact-factorization form.

We presented the first application of F-CT-MQC to the treatment of excited-state processes with explicit inclusion of a time-dependent external field. The algorithm has been adapted to treat periodically-driven systems with the support of the Floquet formalism, and tested on a model systems subject of an external field with different intensities. The results are promising, especially aiming at the combination with quantum-chemistry approaches to compute electronic-structure properties, but there is clearly room for the development of refined approximation strategies to F-CT-MQC that will be the focus of future studies.

\section*{Data availability statement}
The data that support the findings of this study are available from the corresponding author upon reasonable request.


\appendix
\section{Electronic equation of F-CT-MQC}\label{app: el eqn}
The evolution equation for the coefficients $C_{l,m}\big(\mathbf R(t),t\big)$ is derived from Eq.~(\ref{eqn: dot Phi}), using, in the left-hand side, the result of Eq.~(\ref{eqn: Phi dot with Floquet diabatic}), and introducing, in the right-hand side, the expansion of Eq.~(\ref{eqn: Fl dia Phi with R(t)}). In this Appendix, we discuss in detail this procedure, ultimately leading to the derivation of Eq.~(\ref{eqn: dot Csd 3}).

The evolution of the electronic wavefunction~(\ref{eqn: dot Phi}) can be rewritten in the Floquet diabatic basis as
\begin{subequations}\label{eqnapp: Phi dot with Fl}
\begin{align}
\sum_{k,n}\dot C_{k,n}\big(\mathbf R(t),t\big) \psi_k(\mathbf r;\mathbf R(t)) e^{i\omega_nt} =\sum_{k,n}\dot{\mathbf R}(t)\cdot\left(\boldsymbol{\nabla}C_{k,n}\big(\mathbf R(t),t\big)\right)e^{i\omega_nt}\psi_{k}\big(\mathbf r;\mathbf R(t)\big)\label{eqnapp: Phi dot with Fl 1}\\
-\frac{i}{\hbar}\sum_{k,n} \left[\hat H_{el}\big(\mathbf r,\mathbf R(t)\big)+\hbar\omega_n+\hat V(\mathbf r,\mathbf R,t)+\hat U\left[\Phi,\chi\right] \right]C_{k,n}\big(\mathbf R(t),t\big)e^{i\omega_nt}\psi_{k}\big(\mathbf r;\mathbf R(t)\big)\label{eqnapp: Phi dot with Fl 2}\\
+\frac{i}{\hbar}\sum_{k,n} \left[\epsilon\big(\mathbf R(t),t\big)+\epsilon_{\mathrm{ext}}\big(\mathbf R(t),t\big) \right]C_{k,n}\big(\mathbf R(t),t\big)e^{i\omega_nt}\psi_{k}\big(\mathbf r;\mathbf R(t)\big)\label{eqnapp: Phi dot with Fl 3}
\end{align}
\end{subequations}
Here, we rearranged the terms of the original equation, keeping only the term $\dot C_{k,n}\big(\mathbf R(t),t\big)$ in the left-hand side. Moreover, the term containing the spatial derivative $\boldsymbol{\nabla}$ acting on the adiabatic states $\psi_k(\mathbf r;\mathbf R(t))$ in the left-hand side of Eq.~(\ref{eqn: dot Phi}) exactly cancels out the corresponding term on the right-hand side, where the operator $\boldsymbol{\nabla}$ acts on $C_{k,n}\big(\mathbf R(t),t\big)e^{i\omega_nt}\psi_{k}\big(\mathbf r;\mathbf R(t)\big)$. For this reason, only the first term on the right-hand side of Eq.~(\ref{eqnapp: Phi dot with Fl 1}) is left, which contains the spatial derivative of the coefficients (and not of the states).

In order to derive an explicit expression for $\dot C_{k,n}\big(\mathbf R(t),t\big)$ only, from Eqs.~(\ref{eqnapp: Phi dot with Fl}), we project onto a Floquet diabatic state $e^{i\omega_mt}$ $\psi_{l}\big(\mathbf r;\mathbf R(t)\big)$. To this end, as discussed in the main text, we introduce the change a variable $t\rightarrow s$ in the harmonics, and we perform a time integration over a period of the driving field only in the time variable $s$. This operation allows us to integrate out the periodic time dependence due to the external drive and carried by the Floquet diabatic states, while keeping a time-dependent perspective in the ``non-periodic'' variable $t$. The projection in the electronic physical space $\mathbf r$ is performed via the integral over the whole configuration space. In the left-hand side of Eqs.~(\ref{eqnapp: Phi dot with Fl}), the projection in time and space yields
\begin{align}\label{eqnapp: C dot lm}
\dot C_{l,m}\big(\mathbf R(t),t\big)=\frac{1}{T}\int_0^T ds \int d\mathbf re^{-i\omega_ms}\bar\psi_{l}\big(\mathbf r;\mathbf R(t)\big)\sum_{k,n}\dot C_{k,n}\big(\mathbf R(t),t\big) \psi_k(\mathbf r;\mathbf R(t)) e^{i\omega_ns}
\end{align}
where we have used the normalization condition of the Floquet diabatic states
\begin{align}
\frac{1}{T}\int_0^T ds\, e^{i(\omega_n-\omega_m)s}\int d\mathbf r\,\bar\psi_{l}\big(\mathbf r;\mathbf R(t)\big)\psi_k(\mathbf r;\mathbf R(t))  = \delta_{mn}\delta_{lk}
\end{align}
An analogous procedure allows us to derive
\begin{align}
\dot{\mathbf R}(t)\cdot\boldsymbol{\nabla}C_{l,m}\big(\mathbf R(t),t\big)=\sum_{k,n}\frac{1}{T}\int_0^T ds\,e^{i(\omega_n-\omega_m)s} \int d\mathbf r\,\bar\psi_{l}\big(\mathbf r;\mathbf R(t)\big)\dot{\mathbf R}(t)\cdot\left(\boldsymbol{\nabla}C_{k,n}\big(\mathbf R(t),t\big)\right)\psi_{k}\big(\mathbf r;\mathbf R(t)\big)
\end{align}
from the first term on the right-hand side of Eq.~(\ref{eqnapp: Phi dot with Fl 1}), 
\begin{align}
\left[E_l\big(\mathbf R(t)\big)\right.&\left.+\hbar\omega_m\right]C_{l,m}\big(\mathbf R(t),t\big)=\\
&\sum_{k,n}\frac{1}{T}\int_0^T ds\,e^{i(\omega_n-\omega_m)s} \int d\mathbf r\,\bar\psi_{l}\big(\mathbf r;\mathbf R(t)\big) \left[\hat H_{el}\big(\mathbf r,\mathbf R(t)\big)+\hbar\omega_n\right]C_{k,n}\big(\mathbf R(t),t\big)\psi_{k}\big(\mathbf r;\mathbf R(t)\big) \nonumber 
\end{align}
from the first two terms in square brackets in Eq.~(\ref{eqnapp: Phi dot with Fl 2}), and
\begin{align}
\left[\epsilon\big(\mathbf R(t),t\big)\right.&\left.+\epsilon_{\mathrm{ext}}\big(\mathbf R(t),t\big) \right]C_{l,m}\big(\mathbf R(t),t\big) =\\
\sum_{k,n} \frac{1}{T}&\int_0^T ds\,e^{i(\omega_n-\omega_m)s} \int d\mathbf r\,\bar\psi_{l}\big(\mathbf r;\mathbf R(t)\big)\left[\epsilon\big(\mathbf R(t),t\big)+\epsilon_{\mathrm{ext}}\big(\mathbf R(t),t\big) \right]C_{k,n}\big(\mathbf R(t),t\big)\psi_{k}\big(\mathbf r;\mathbf R(t)\big)\nonumber 
\end{align}
from Eq.~(\ref{eqnapp: Phi dot with Fl 3}). In Eq.~(\ref{eqnapp: Phi dot with Fl 2}), the term containing the external drive is projected onto the Floquet diabatic state $e^{i\omega_mt}\psi_{l}\big(\mathbf r;\mathbf R(t)\big)$, by performing the time-variable transformation $t\rightarrow s$ in $\hat V(\mathbf r,\mathbf R(t),t)$, before the integration over a period, such that
\begin{align}
\sum_{k,n}V_{mn,lk}&\big(\mathbf R(t)\big)C_{k,n}\big(\mathbf R(t),t\big)= \nonumber\\
&\sum_{k,n}\frac{1}{T}\int_0^T ds\,e^{i(\omega_n-\omega_m)s} \int d\mathbf r\,\bar\psi_{l}\big(\mathbf r;\mathbf R(t)\big)\hat V(\mathbf r,\mathbf R(t),s)\psi_{k}\big(\mathbf r;\mathbf R(t)\big)C_{k,n}\big(\mathbf R(t),t\big)
\end{align}
The expression of the electron-nuclear coupling operator $\hat U\left[\Phi,\chi\right]$ used in Eq.~(\ref{eqnapp: Phi dot with Fl 2}) is obtained by representing the nuclear wavefunction in terms of its modulus and phase, such that 
\begin{subequations}
\begin{align}
\hat U[\Phi,\chi] &\simeq\mathbf M^{-1} \left(\boldsymbol{\nabla} S\big(\mathbf R(t),t\big)+\mathbf A\big(\mathbf R(t),t\big) +i \frac{-\hbar\boldsymbol{\nabla}\left|\chi\big(\mathbf R(t),t\big)\right|^2}{\left|\chi\big(\mathbf R(t),t\big)\right|^2}\right)\cdot
\left(-i\hbar\boldsymbol{\nabla}-\mathbf A\big(\mathbf R(t),t\big)\right)\label{eqn: Uen with R(t)}\\
&= \left[\dot{\mathbf R}(t)+i\mathbf M^{-1}\boldsymbol{\mathcal P}\big(\mathbf R(t),t\big)\right]\cdot\left(-i\hbar\boldsymbol{\nabla}-\mathbf A\big(\mathbf R(t),t\big)\right).\label{eqn: Uen with R(t) and Q(t)}
\end{align}
\end{subequations}
The first term in parenthesis of Eq.~(\ref{eqn: Uen with R(t)}) is the velocity of the trajectory, $\dot{\mathbf R}(t)$, from the characteristic equation~(\ref{eqn: R dot}), whereas the second term contains the quantum momentum~\cite{Gross_PRL2015}, $\boldsymbol{\mathcal P}(\mathbf R(t),t)$, that has been defined in previous work on CT-MQC (but it has a longer history in the context of the quantum-trajectory formalism~\cite{Rassolov_JCP2004, Rassolov_CPL2003}). The quantum momentum induces quantum decoherence effects by tracking the spatial delocalization over time of the nuclear density (or, equivalently, of its modulus)~\cite{Agostini_EPJB2018, Maitra_JCTC2018}. Note that the definition~(\ref{eqn: enco}) of $\hat U[\Phi,\chi]$ contains two terms, but the first term has been shown~\cite{AgostiniEich_JCP2016, Scherrer_PRX2017, Schild_JPCA2016, Scherrer_JCP2015} to be smaller if compared to the second term, and we will thus neglect it henceforth. In addition, such term contains second-order spatial derivatives of Floquet diabatic states, which are usually not available in quantum-chemistry codes to compute electronic-structure properties. The corresponding term in the TDPES~(\ref{eqn: TDPES}) is neglected as well to maintain gauge invariance of the exact-factorization equations in the trajectory-based formulation~\cite{Gross_PRL2015, Gross_JCTC2016}.

Evaluating the term $-i\hbar\boldsymbol{\nabla}-\mathbf A\big(\mathbf R(t),t\big)$ of Eq.~(\ref{eqn: Uen with R(t) and Q(t)}) for the remaining part of Eq.~(\ref{eqnapp: Phi dot with Fl 2}) that contains the quantum momentum, we get
\begin{align}\label{eqnapp: term with qmom}
-i\hbar\boldsymbol{\nabla}&C_{l,m}\big(\mathbf R(t),t\big)-\mathbf A\big(\mathbf R(t),t\big)C_{l,m}\big(\mathbf R(t),t\big) -i\hbar\sum_{k,n} C_{k,n}\big(\mathbf R(t),t\big)\mathbf d_{lk}\big(\mathbf R(t)\big)\delta_{mn} =\\
&\sum_{k,n}\frac{1}{T}\int_0^T ds\,e^{i(\omega_n-\omega_m)s} \int d\mathbf r\,\bar\psi_{l}\big(\mathbf r;\mathbf R(t)\big)\left(-i\hbar\boldsymbol{\nabla}-\mathbf A\big(\mathbf R(t),t\big)\right)C_{k,n}\big(\mathbf R(t),t\big)\psi_{k}\big(\mathbf r;\mathbf R(t)\big)\nonumber
\end{align}
Here, the nonadiabatic coupling vectors appear, namely
\begin{align}
\mathbf d_{lk}\big(\mathbf R(t)\big)\delta_{mn} =\frac{1}{T}\int_0^Tds\,e^{i(\omega_n-\omega_m)s} \int d\mathbf r\,\bar\psi_{l}\big(\mathbf r;\mathbf R(t)\big)\boldsymbol{\nabla}\psi_{k}\big(\mathbf r;\mathbf R(t)\big)
\end{align}
that only contribute to population transfer between states $k$ and $l$ that belong to the same harmonic, due to the presence of $\delta_{mn}$ in the first line of Eq.~(\ref{eqnapp: term with qmom}).

Putting together the results from Eq.~(\ref{eqnapp: C dot lm}) to Eq.~(\ref{eqnapp: term with qmom}), we get the time evolution of the expansion coefficients of the electronic wavefunction in the Floquet diabatic basis along the trajectory $\nu$
\begin{subequations}
\begin{align}
\dot C_{l,m}^\nu(t)&=-\frac{i}{\hbar} \left[E_l^\nu+\hbar\omega_m-\epsilon_{\mathrm{ext}}^\nu(t)-\left(\epsilon^\nu(t)+\dot{\mathbf R}^\nu(t)\cdot\mathbf A^\nu(t)\right)\right]C_{l,m}^\nu(t)\\
&+\frac{\mathbf M^{-1}\boldsymbol{\mathcal P}^\nu(t)}{\hbar}\cdot\left(-i\hbar\boldsymbol{\nabla}C_{l,m}(t)^\nu-\mathbf A^\nu(t) C_{l,m}^\nu(t)\right)\\
&-\sum_{k,n} \left(\dot{\mathbf R}^\nu(t)\cdot\mathbf d_{lk}^\nu\delta_{mn}C_{k,n}(t)
-\left[\mathbf M^{-1}\boldsymbol{\mathcal P}^\nu(t)\right]\cdot\mathbf d_{lk}^\nu\delta_{mn}C_{k,n}^\nu(t)
-\frac{i}{\hbar}V_{mn,lk}^\nu C_{k,n}^\nu(t)\right)
\end{align}
\end{subequations}
Few additional steps are necessary to arrive at the final expression given in Eqs.~(\ref{eqn: dot Csd 3}), namely:\\
(i) we set the gauge freedom such that $\epsilon^\nu(t)+\dot{\mathbf R}^\nu(t)\cdot\mathbf A^\nu(t)=0$ (see discussion in Section~\ref{sec: F-CT-MQC});\\
(ii) we neglect the term containing the product quantum momentum $\boldsymbol{\mathcal P}^\nu(t)$ and of the nonadiabatic couplings $\mathbf d_{lk}^\nu$ (see Refs.~\cite{Agostini_JCTC2020_1, Gross_JPCL2017, Gross_JCTC2016});\\
(iii) we approximate $\boldsymbol{\nabla}C_{l,m}^\nu(t)$ such that we neglect the spatial dependence of the modulus of $C_{l,m}^\nu(t)$ and we only keep the spatial derivative of the phase, which we call $\mathbf f_{l,m}^\nu $, i.e., $-i\hbar\boldsymbol{\nabla}C_{l,m}^\nu \simeq C_{l,m}^\nu\mathbf f_{l,m}^\nu $ (see Refs.~\cite{Agostini_JCTC2020_1, Gross_JCP2015, Gross_PRL2013}).

These additional approximations allow us to derive Eq.~(\ref{eqn: dot Csd 3}).
%


%

\end{document}